\documentclass[authoryear,11pt]{elsarticle}
\usepackage[margin=0.80in]{geometry}
\usepackage{float}
\usepackage{adjustbox}
\usepackage{wrapfig}
\usepackage{soul}
\usepackage{footnote}
\usepackage{verbatim}

\usepackage{multirow}
\usepackage{graphics}
\usepackage[dvips]{color}
\usepackage{pifont}
\usepackage{graphicx}

\usepackage[font=small,skip=0pt]{caption}
\usepackage{subcaption}

\usepackage{amssymb,amstext,amsmath}
\usepackage{float}
\usepackage{booktabs}
\usepackage{array,ragged2e}
\usepackage[table]{xcolor}
\usepackage{multirow}
\usepackage{algorithm}
\usepackage{algpseudocode}
\usepackage{pifont}

\journal{}
\begin{document}
\begin{frontmatter}

\title{Assessment of System-Level Cyber Attack Vulnerability for Connected and Autonomous Vehicles Using Bayesian Networks}

\author[PEaddress,UIUCaddress]{Gurcan Comert\corref{mycorrespondingauthor}}
\cortext[mycorrespondingauthor]{Corresponding author}
\ead{gurcan.comert@benedict.edu}
\author[CUaddress]{Mashrur Chowdhury} 
\ead{mac@clemson.edu}
\author[UIUCaddress]{David M. Nicol}
\ead{dmnicol@illinois.edu}

\address[PEaddress]{Computer Science, Physics, and Engineering Department, Benedict College, 1600 Harden St., Columbia, SC 29204 USA}
\address[CUaddress]{Glenn Department of Civil Engineering, Clemson University, 216 Lowry Hall, Clemson, SC 29634 USA}
\address[UIUCaddress]{Information Trust Institute, University of Illinois, Urbana-Champaign, IL 61801 USA
\vspace{-10mm}}

\begin{abstract}
This study presents a methodology to quantify vulnerability of cyber attacks and their impacts based on probabilistic graphical models for intelligent transportation systems under connected and autonomous vehicles framework. Cyber attack vulnerabilities from various types and their impacts are calculated for intelligent signals and cooperative adaptive cruise control (CACC) applications based on the selected performance measures. Numerical examples are given that show impact of vulnerabilities in terms of average intersection queue lengths, number of stops, average speed, and delays. At a signalized network with and without redundant systems, vulnerability can increase average queues and delays by $3\%$ and $15\%$ and $4\%$ and $17\%$, respectively. For CACC application, impact levels reach to $50\%$ delay difference on average when low amount of speed information is perturbed. When significantly different speed characteristics are inserted by an attacker, delay difference increases beyond $100\%$ of normal traffic conditions.
\end{abstract}
\end{frontmatter}
\section{Introduction}
Transportation networks are one of the critical infrastructures that need to be strategically managed via vulnerabilities for dynamic threats and hazards to achieve secure and resilient cities (\cite{cardenas2011attacks,chowdhury2019security}). Resilience includes a protective strategy against unknown or highly uncertain events. Possible tools to improve resiliency can be listed as diversification and sensors, reducing vulnerability, design of flexible systems, and improve conditions for system adaptation (\cite{aven2013uncertainty}). In intelligent transportation systems (ITS), in order to increase security and resiliency (\textit{i.e.}, include safety and mobility) against possible attacks or benign system errors by different events (\cite{mitchell2014survey},\cite{humayed2017cyber}), research is needed on physical models for attack, detection, and mitigation using various redundant sensors and additional information. Furthermore, level of flexibility and predictability also need be investigated research for control systems. Such system would have to keep conventional sensors for redundancy as well as validating reading from connected components (\cite{peisert2014designed}). 

Intelligent transportation systems applications with connected and autonomous vehicles (CAVs) assume some market penetration level and rely on them for system control such as traffic signals, emergency vehicle management, cooperative adaptive cruise control (CACC) or platooning, and real-time routing etc. Thus, any additional uncertainty would cause sub-optimal system control, failures, and incidents in safety-critical applications. In this paper, we aim to address this issue with system level uncertainty quantification of cyber attacks on ITS applications particularly involving CAVs and methods to incorporate such in stochastic models based on the National Intelligent Transportation System Reference Architecture (ARC-IT) and the Common Vulnerability Scoring System (CVSS). Attack types on CAVs can be listed but not limited to jamming, spoofing, illusion, blinding, replay, data manipulation, eavesdropping, routing, DOS, impersonation, and coordinated multiple attacks (\cite{petit2015potential,chowdhury2019security}). Author's previous work looked at simpler version for isolated intelligent signals (\cite{comert2018modeling}). This study significantly extends previous work by incorporating cooperative adaptive cruise control (CACC or also called platooning) with microsimulations and network version of both BNs and intelligent signals. Approach is modular so model can be improved when detection interfaces and series of other vehicles added. More interestingly, series of vehicles in a platoon (\textit{i.e.}, probabilistic graph model as hidden Markov model) are given to propagate risk in multiple interacting networks (i.e., under connected vehicles). 

Attack detection and mitigation on physical system models are some of the recent approaches gaining interest. For different types of cyber attacks (\textit{e.g.}, (\cite{petit2015potential,van2016survey})), risk areas (components or nodes), cost of different attacks, and detection of abnormal messages within driver, vehicle, and infrastructure context are modeled via probabilistic graphical models (e.g., Bayesian Networks (BNs) (\cite{al2013context,cao2015preemptive})) and traffic simulations with various scenarios. Results would be used for strategic allocation of redundancy systems, optimal detection deployment, and generation of mitigation techniques. The critical issue, however, when data and interaction among CAVs included, any error would result in significant costs. For instance, false revoking emergency vehicle security certificate would result in life-endangering situations. Similarly, delay caused by device flooding or denial of service (DOS) on transit and freight trucks would result as excessive delay, fuel consumption, and emissions. 

Vulnerability defined as combination of the impacts of the application considered and associated uncertainties (levels) given an initiating event (risk source/attack surface) (\cite{aven2015risk}) and formulated as Vulnerability $|$ $A$=Impacts$+$Uncertainty $|$ the occurrence of the event $A$. Risk quantification is one of the fundamental research areas under decision support systems. One of the key review studies summarizes the research and the needs in last decade (\cite{aven2016risk}). As targeted by this study, risk assessment on interdependent critical infrastructures was listed as one of the needs. The author also emphasized developing unifying platform related to risk, vulnerability, and probability. In a relevant study, attack impact cost was quantified by binomial lattice model (\cite{khansa2009valuing}) on firms' information security investments. The study provides switching options in case of emergency. 

Attack modeling in game-theoretic framework is also gaining interest (\cite{cheung2019attacker}). However, realistic physical model of the problem is critical recognized by the researchers similar to industrial control systems (\cite{sridhar2014model,li2019multistream,maziku2019security}). For instance on flood control system, risk and impact were quantified by factorial probabilistic inference in (\cite{wang2016risk}). In another fundamental review study (\cite{mitchell2014survey}), authors studied reputation management in vehicular networks for both detection and mitigation. Likewise, in (\cite{van2016survey}) for ITS applications, revoking or blacklisting the information producers are presented. One of two main issues in CAVs, safety critical applications and privacy of the drivers are also investigated by the researchers (\cite{sucasas2016autonomous}). In order to facilitate easier and cost effective decision making, standardization of actions based on index calculated for various attacker type (i.e., intelligent or random) were attempted also (\cite{bier2019risk}). Cost of impact of broad cyber events in firms was modeled using extreme event theory (\cite{eling2019actual}). Similar to this study, modular and probabilistic approach allows modelers to be able to incorporate other attack types and their likelihoods easily.
\subsection{Contribution of this paper}
In this study, traffic systems are modeled analytically via a probabilistic graphical networks to demonstrate possible vulnerabilities, propagation of such vulnerabilities, impacts of them as performance measures, and effect of possible redundant components on the vulnerabilities are developed for decision makers. Particular contributions can be listed as: (1) uncertainty quantification on ITS applications with cyber-physical representation, (2) application of a probabilistic expert systems (BNs) for modeling anomalies and attacks (malicious messages/benign failures) are produced, and (3) impact of redundancy systems (sensors and surveillance systems) and number of different traffic states (versus control reliability) are incorporated.
\subsection{Organization of the paper}
The paper is organized as follows. Section~\ref{sctmethod} discusses the approach's relevance in risk analysis and decision support. Section~\ref{sctpd} introduces the problem and the approach. Section~\ref{sctbn} explains the modeling and Bayesian networks.  Section~\ref{sctsims} presents simulations from BNs and process simulators and their results. Section~\ref{sctne} includes the detailed numerical experiments for quantification of uncertainty due to attacks on CACC (or platooning) and traffic network with intelligent signals (ISIG). The section also describes analysis with and without sensors at isolated and network of signals. Finally, section~\ref{sctconc} summarizes findings and addresses possible future research directions. 
\section{Approach}
\label{sctmethod}
\begin{figure*}[h!]
\centering
  \fbox{\includegraphics[width=0.90\linewidth]{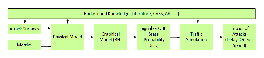}}
\caption{Flow diagram of the methodology for vulnerability $V$=$(C',Q,K|A)$ assessment (based on \cite{aven2011misconceptions})}
  \label{fig_method}
\end{figure*}

Throughout the paper, fundamental definitions were adopted from \cite{aven2015risk}. Vulnerability is considered to be part of risk analysis and risk analysis can be referred as risk and vulnerability analysis \cite{aven2015risk}. Concept of vulnerability is simply defined as risks given occurrence of an event, (\textit{i.e.},Risks$|$Threats$\lor$Hazard$\lor$Attack). A conditional event can be described as $C'$ factors to define consequences or impacts, uncertainty $Q$ as probabilities, and $K$ background knowledge from cited works. Thus, vulnerability of system can be expressed as $V$=$(C',Q,K|A)$. This relation summarizes our approach presented in Fig.~\ref{fig_method}. 

Overall, our approach falls under operational vulnerability, vulnerability analysis+ evaluation equal to vulnerability assessment (\cite{aven2013uncertainty}). And, assessment and some steps from planning are covered in the risk analysis process. Vulnerability management consists of high risk+uncertainties leads challenge in this difficulty in predicting outcomes. By \cite{aven2013uncertainty}, benefits of a vulnerability analysis are listed as: (1) drawing a vulnerability picture, (2) comparing alternative solutions with respect to vulnerability, (3) identifying factors and components that are critical with respect to vulnerability, (4) demonstrating the effect of various measures to vulnerability, and (5) conducting this analysis mainly to support decision making concerns, safety, and costs. Uncertainty in vulnerability assessment can be expressed as identification of relevant threats, hazards, cause, and consequences analysis, including assessment of exposures and vulnerable risk description. 

Similarly, the basis for the analysis in the paper: (1) evaluations of ITS applications by design, (2) drawing conclusions on what design and measure meet the stated requirements in current standards, (3)  evaluating related performance of the preparedness of ITS systems, and (4) documenting the vulnerability levels. Hence, generate knowledge to support decision making that would provide an important basis for finding the right balance among different hazards, components, safety, and costs (\cite{aven2013uncertainty}).
\section{Problem Definition}
\label{sctpd}
\begin{figure*}[h!]
\centering
\begin{subfigure}{.98\textwidth}
\centering
  \fbox{\includegraphics[width=0.83\linewidth]{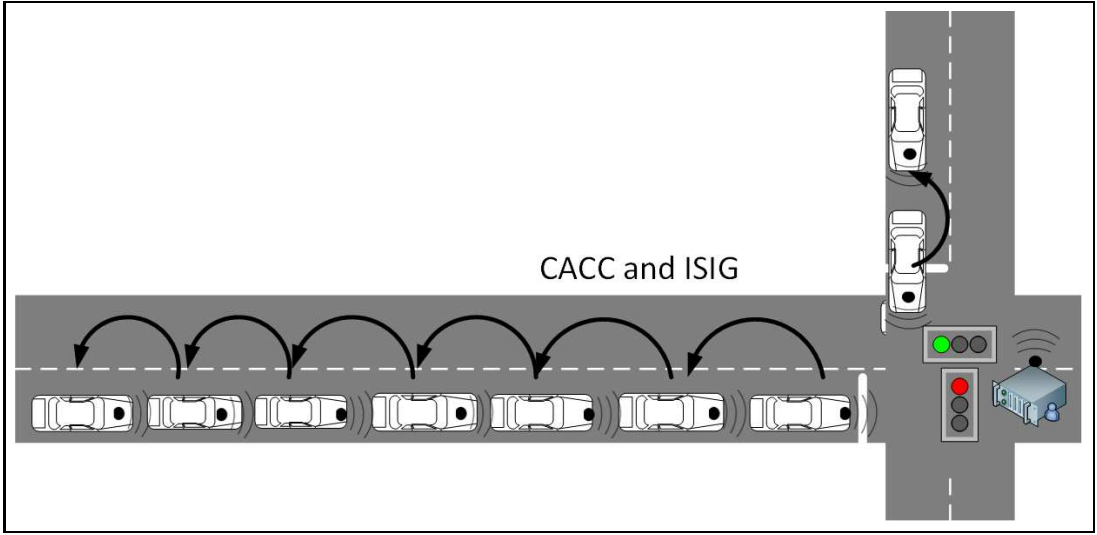}}
\caption{Example ITS CACC and ISIG applications}
  \label{fig_intersection}
\end{subfigure}%
\\
\begin{subfigure}{.98\textwidth}
 \centering
\fbox{\includegraphics[width=0.83\linewidth]{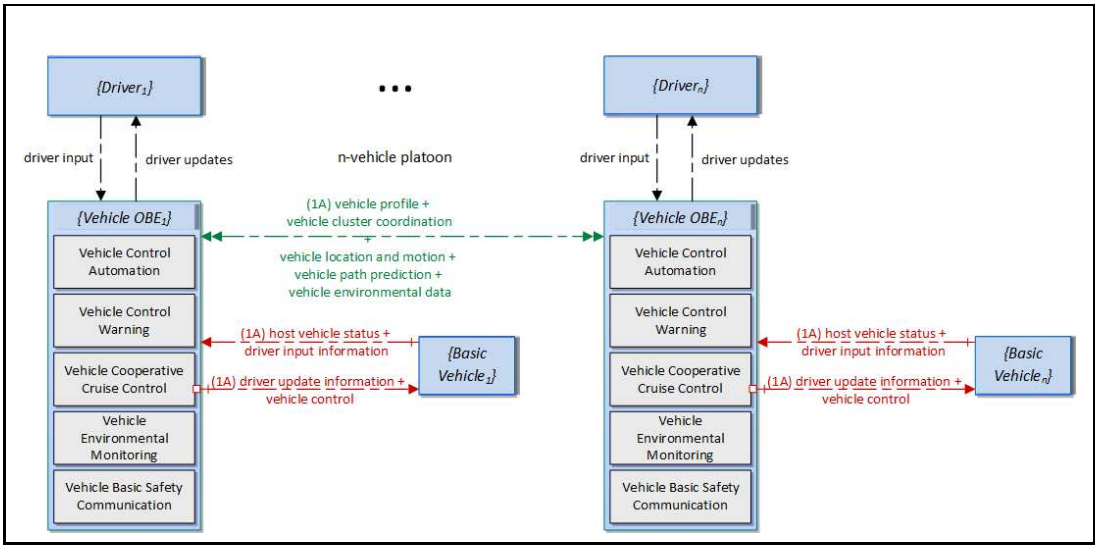}}
  \caption{A possible CACC application (ARC-IT Physical Diagram) adopted from \cite{arcit}}
  \label{fig_cacc}
\end{subfigure}
\caption{Methodology}
\label{fig_phys}
\end{figure*}

Primary goal in this paper is to develop models for probabilistic vulnerabilities (\cite{borgonovo2018risk}) in interested ITS applications with respect to considered attack surfaces ($A$) (see Fig.~\ref{fig_method}) and quantify the impact on traffic flow. Within vulnerability $(C',Q,K|A)$ assessment framework, as graphical model, Bayesian networks are modeled based on physical diagram of applications (\textit{e.g.}, CACC and ISIG in Fig.~\ref{fig_intersection}) defined in \cite{arcit} (e.g., knowledge $K$). In the second step, risk levels are obtained from (\cite{petit2015potential, comert2018modeling}) (also $K$). Third, vulnerability scores are calculated from propagated conditional probabilities ($Q$) based on (\cite{cvss}). Lastly, simulations are given for impact ($C'$) studies to quantify propagated risk on traffic.

Based on uncertainty assessment of vulnerabilities (\cite{aven2013uncertainty}), \textit{i.e.}, process to classify the strength of metrics discussed by the authors, studied cases can be categorized as medium strength of knowledge to be able to infer exact and critical assignment. However, variables, state values, and levels are adopted from \cite{arcit} and \cite{cvss}.

In Fig.~\ref{fig_cacc}, possible CACC components are shown which represent physical devices. In a platoon (with autonomous vehicles) setting, each vehicle can have a communication device (i.e., onboard equipment ($OBE$)) (\cite{dey2015review}). For the nodes, the security certificate authority (e.g. USDOT-security credential management system ($SA$)) is interfaced to OBEs and is used to provide security certificates to trusted OBEs. Some or all of these vehicles can be equipped, that is, they have some type of portable device which can communicate traffic control system (via $RSE$). Traffic control system consists of signal controller, other surveillance technologies and environmental sensors, and the RSE (or a communication device). The RSE radio referred to the $5.9$ GHz Dedicated Short Range Communication (DSRC) communications between the vehicles and the infrastructure, however, other wireless communications can also be considered. For communications, specific channels are designated for broadcasting basic safety messages ($BSM$s), the map message, and the signal phase and timing messages, and signal status messages. 

Rather broader view of a traffic network, controllers can be viewed as a part of traffic management system ($TMC$) as shown in Fig~\ref{fig_cacc}. Thus, different control devices can be connected via communications can be coordinated for various modes of transportation. These modes can be sensed by different detection technologies (fixed-inductive loops, video, infrared, radar, magnetic plates, pedestrian buttons or solely mobile-GPS, and cell phones etc.). These detectors provide information to the control algorithms for presence to change application states. Some of these communications are direct (pedestrian push button) and some are based on inference (video cameras-queues). 

Functions of the OBE in a CACC are processing other OBE data, processing pedestrian protection, processing indicator output data for roads, monitoring RSE operation, processing indicator output data for signals or users, and provide device interface for field management stations. Impact on applications can be listed such as data distribution for speed harmonization, emergency vehicle preemption, freight signal priority, pedestrian mobility, and red light violation warning. For simplification, however, impact of attacks on platooning is represented as changes in exchanged speed by platoon leader's OBE, i.e., average and total delays.
 
In \cite{comert2018modeling}, possible intelligent signal components was given. An intelligent signal (ISIG) as a system would include a controller unit with a processor and connections (\textit{e.g.}, ethernet or wireless, or other ports) (see also architecture of connected adaptive signal (\cite{MMITSS2016})). This physical signal model is utilized to test possible attack scenarios and impacts. In the network setting, each intersection is assumed to be equipped with a communication technology. A traffic management system is also included. 

Functions of an intelligent signal controller can be listed as processing: traffic sensor data; pedestrian protection; indicator output data for roads and for other signals or users; RSE operation and signal control conflict monitoring; and device interface for field management stations. Impact on applications can be experienced through data distribution for eco-signaling, eco-transit, emergency vehicle preemption, freight signal priority, pedestrian mobility, red light violation warning, transit signal priority. In this paper, impact of attacks on ISIGs is represented as changes in service times thus average and maximum queue lengths in vehicles and delays in seconds (time spent in queue and server).   
\section{Bayesian Networks}
\label{sctbn}
Bayesian networks are directed acyclic graphical stochastic expert systems that depict conditional (in)dependence of probability distributions of interests. In this section, Bayesian or Belief networks for system risk analysis is discussed (\cite{rausand2013risk,aven2015risk}). Probabilistic graphical models represent the causal connections among various variables denoted as nodes at different levels (\cite{cai2011identifying}). These nodes can be at many states. Hence, more complex systems can be analyzed by BNs compared to fault tree analysis with two states. In order to determine conditional probabilities given causal connectivity direct, argument, or specified procedures, BNs constructed that establish the probabilities for the various quantities (events) in the network. Then, the probabilities are calculated. 
BNs in cybersecurity context are utilized to model hypothesized communications and generate vulnerability/reliability of the network tested with possible attacks at traffic networks (\cite{al2013context, huang2014evidence, wan2014context}). Example cases on safety can be seen in \cite{mackay2003information, barber2012bayesian} and other applications in \cite{chin2009assessing,zhang2013decision,gill2019behavior}. 

In this study, such hazards are adapted for CAVs framework (\cite{petit2015potential}). Physical model in Fig.~\ref{fig_cacc} are modified via addition of detection interfaces. Additionally, intelligent signal model was included multiple modes, back up surveillance systems, and series of other signals in BNs and simulations. BNs are employed to calculate probability distribution, expected vulnerability and its standard deviation via Monte Carlo simulations, and some predictions are provided given assumed evidence for different scenarios. Variables and their discrete states are defined. Note that these are assumptions mostly compiled from (\cite{petit2015potential,arcit,cvss}). First, nodes and edges are defined from the Fig.~\ref{fig_cacc}. These nodes are the variables that are impacting vehicle on-board equipment ($OBE$) state. Parental nodes along with their vulnerability scores are given as in Table~\ref{tab_surf} (\cite{arcit,petit2015potential,comert2018modeling}).

Table~\ref{tab_surf} essentially is a risk matrix with description based on consequence categories. Again vulnerability is given an threat or event $V$=$Risk|A$, even after simulations how can we model this vulnerability is needed. Step-by-step, the expected value of $E(C'|A)$ with states of level of impacts (consequences) such as $low$, $medium$, and $high$ are deduced. This expectation can be calculated by $P(A)E(C'|A)$, \textit{i.e.}, Vulnerability Score=P(attack occurs)P(attack result in damage attack occurs)E(impact$|$attack and results in impact) (\cite{aven2015risk}). This main approach is risk analysis method of classification based on expected consequences. 

Regardless, assumed probability distributions may not adequately explain the event likelihoods then vulnerabilities. In an interesting discussion (\cite{aven2012foundations}), possibility theory deals with uncertainty character in the case of incomplete information. Thus, different probability and necessity measures are assigned for risk. This approach is closely related fuzzy logic or Grey systems. Evidence (Dempster-Shafer) theory with belief functions is very similar to Bayesian treatment. In our approach, a probabilistic vulnerability assessment with uncertainty propagation is adopted with initial probabilities and evidence-Bayesian from existing literature knowledge.    
\begin{table}[h!]
\centering
\caption{Attack surfaces for equipped vehicles and pedestrians}
\label{tab_surf}
\scalebox{0.75}{
\begin{tabular}{llllll}
  \hline
  Target & Type & Feasibility & Metric & Detection & Metric \\ 
  \hline
 SA & LTC & low & $[h\ l\ l]$  & med & $[l\ h\ l]^T \times [h\ l\ l]$ \\ 
   & CRL & med & $[l\ h\ l]$ & med & $[l\ h\ l]^T \times[l\ h\ l]$ \\ 
    & PC & med & $[l\ h\ l]$ & med & $[l\ h\ l]^T \times[l\ h\ l]$ \\ 
   RSE & WSA & high & $[l\ l\ h]$ & low & $[h\ l\ l]^T \times[l\ l\ h]$ \\ 
    & DB & high & $[l\ l\ h]$ & med & $[l\ h\ l]^T \times[l\ l\ h]$ \\ 
    & DOSR & high & $[l\ l\ h]$ & high & $[l\ l\ h]^T \times[l\ l\ h]$ \\ 
    & SD & low & $[h\ l\ l]$ & high & $[l\ l\ h]^T \times[h\ l\ l]$ \\ 
   VEHs & CB & low & $[h\ l\ l]$ & low & $[h\ l\ l]^T \times[l\ h\ l]$ \\ 
    & BLK & med & $[l\ h\ l]$ & low & $[h\ l\ l]^T \times[l\ h\ l]$ \\ 
    & RB & low & $[h\ l\ l]$ & low & $[h\ l\ l]^T \times[h\ l\ l]$ \\ 
    & BSM & high & $[l\ l\ h]$ & low & $[h\ l\ l]^T \times[l\ l\ h]$ \\ 
    & DOS & high & $[l\ l\ h]$ & low & $[h\ l\ l]^T \times[l\ l\ h]$ \\ 
    & MP & high & $[l\ l\ h]$ & med-high & $[l\ m\ h]^T \times[l\ l\ h]$ \\ 
    & DCC & med & $[l\ h\ l]$ & med & $[h\ l\ l]^T \times[l\ h\ l]$ \\ 
    & LC & med & $[l\ h\ l]$ & low & $[h\ l\ l]^T \times[h\ l\ l]$ \\ 
   \hline
\end{tabular}
}
\end{table}
\subsection{CACC Application}
For a CACC application model, constraints of the system or assumed variables, their description, and metrics (states) are given in Table~\ref{tab_metrics2}. As for intelligent signals, attack surfaces are adopted from the list in Table~\ref{tab_surf}. 
\begin{table}[h!]
\centering
      \caption{Nodes on the designed BNs with assumed metrics for CACC}
      \label{tab_metrics2}
      \centering
\scalebox{0.62}{
\begin{tabular}{lllll}
  \hline
  Node & Description & Confidentiality & Integrity & Availability \\ 
  \hline
   RSE & connected vehicle roadside equipment & M & M & M \\ 
   ITS & ITS roadway equipment & M & H & M \\ 
   TMC & traffic management center & M & H & M \\ 
   OBE & vehicle OBE & L & M & M \\ 
   BV & basic vehicle & L & M & L \\ 
   OV & other vehicle OBEs & L & H & M \\ 
   DII & driver input information & M & H & H \\ 
   DI & driver input & M & H & H \\ 
   VCC & vehicle cluster coordination & L & H & M \\ 
   VVV & vehicle path prediction & NA & H & M \\ 
   VP & vehicle profile & L & M & M \\ 
   DRV & driver update information & L & M & M \\ 
   VCA & vehicle control automation & M & H & H \\ 
   DR & driver updates & NA & M & M \\ 
   BSC & basic safety communication & L & M & M \\ 
   HVS & host vehicle status & L & M & H \\ 
   DI & driver input & M & H & H \\ 
   VED & vehicle environmental data & L & M & M \\ 
   VLM & vehicle location and motion & L & H & M \\ 
   VCW & vehicle control warning & L & H & H \\ 
   VCP & vehicle cooperative cruise control & L & H & M \\ 
   BSM & basic safety message & L & L & H \\ 
   DD & detection by driver & L & M & H \\ 
   DS & detection by system & L & M & H \\ 
   RDD & roadside detection by system driver & L & M & H \\ 
   RDS & roadside detection by system & L & M & H \\
   SA & certificate (security) authority & L & M & H \\ 
   WSA & wrong safety warning and signal phasing & L & M & H \\ 
   CB & sending channel busy & L & M & H \\ 
   DCC & distribution congestion conrol mechanism & L & M & H \\ 
   LC & location tracking & L & M & H \\ 
   RDS & detection by system  & L & M & H \\ 
   LTC & fake long term certificate & L & M & H \\ 
   CRL & fake certificate revocation list & L & M & H \\ 
   PC & pseudonym certificate & L & M & H \\ 
   DB & database map poisining & L & M & H \\ 
   MP & map poisining (Web) & L & M & H \\ 
   PC & pseudonym certificate & L & M & H \\  
   DOSR & denial of service (RSE) & L & M & H \\ 
   SD & device shutdown & L & M & H \\  
   BLK & block pseudonym change  & L & M & H \\ 
   RB & remote reboot & L & M & H \\  
   DOS & denial of service (vehs,pedes) & L & M & H \\    
   \hline
\end{tabular}
}
\end{table}

\begin{figure*}[h!]
 \centering
 \begin{subfigure}{.44\textwidth}
  \centering
 \fbox{\includegraphics[width=0.988\linewidth]{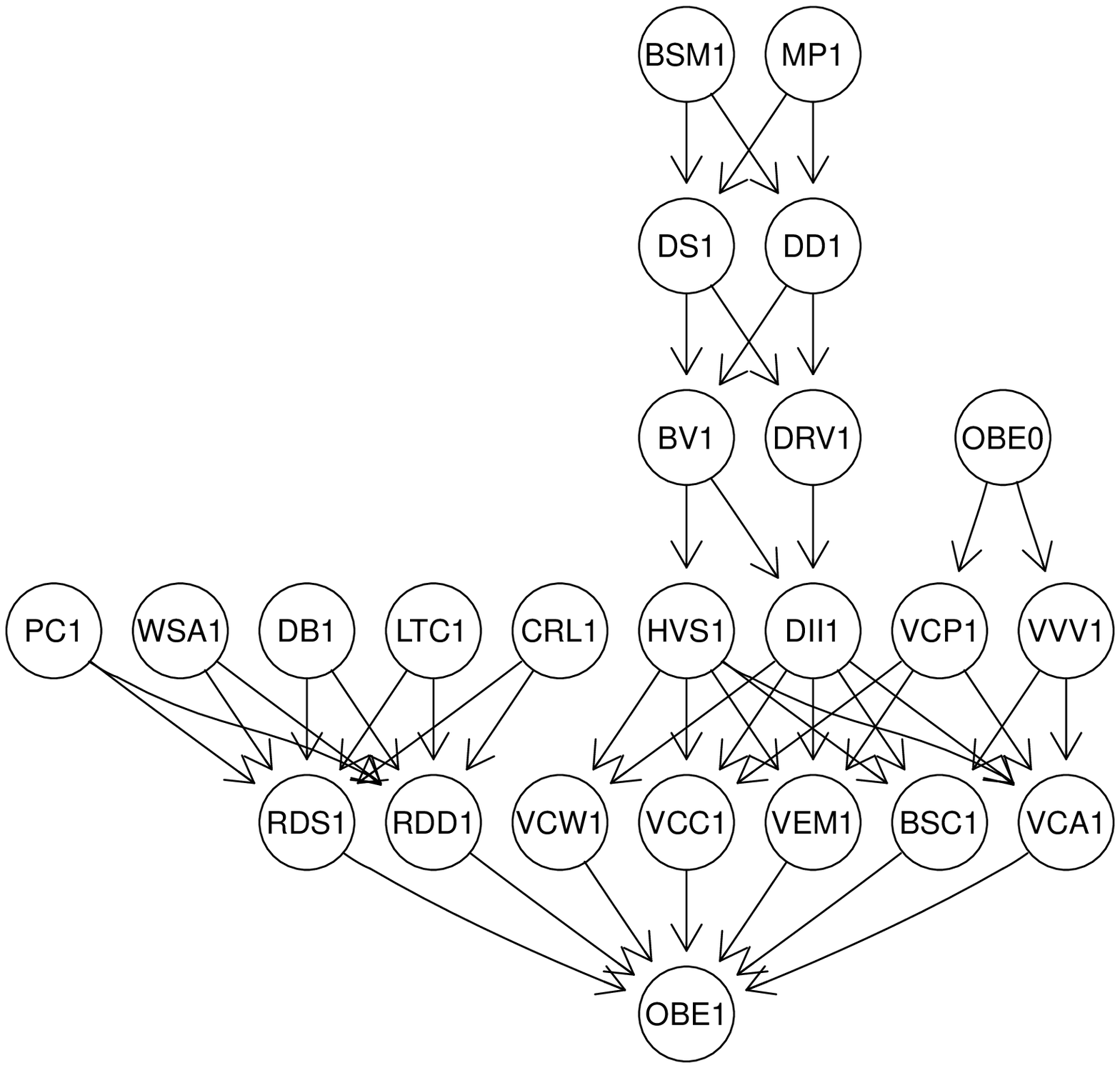}}
   \caption{BN with single OBE}
   \label{fig_bn1}
 \end{subfigure}
 \begin{subfigure}{.4478\textwidth}
 \centering
   \fbox{\includegraphics[width=0.988\linewidth]{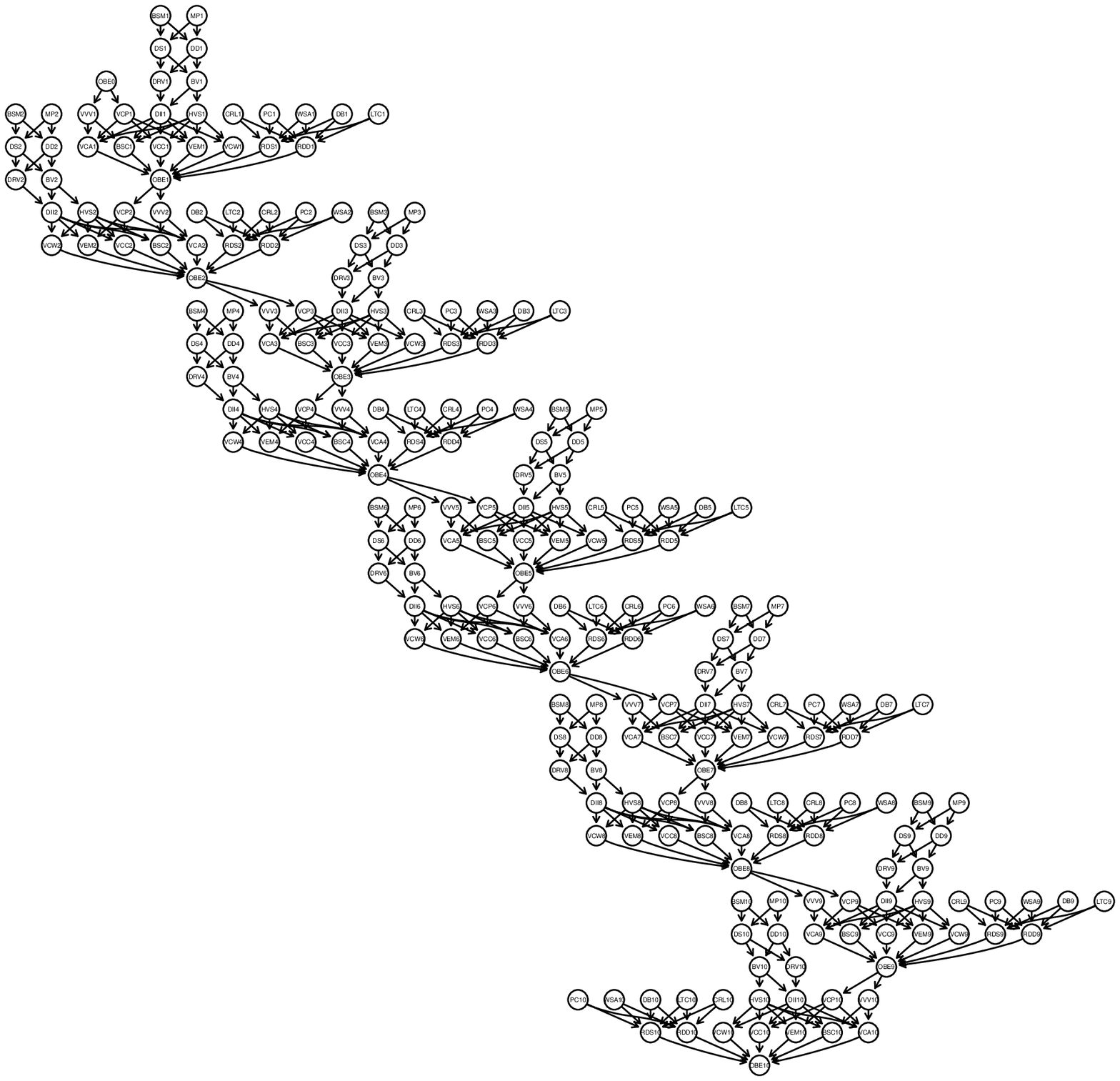}}
 \caption{BN with 10 OBEs}
   \label{fig_bn10}
 \end{subfigure}
  \caption{BNs developed for CACC of Fig. \ref{fig_cacc}}
 \label{fig_bn}
 \end{figure*}

Fig.~\ref{fig_bn1} and \ref{fig_bn10} show the designed BNs under dependencies among the nodes given in Table~\ref{tab_metrics2} for single $OBE$ and a platoon with $10$ vehicles connected through $OBE$s, respectively. This is critical as reliability of $OBE$ could increase via additional ground truth check or observer. However, it will need additional cost. Similarly, detection nodes can be seen as cost for monitoring and delay of communications. In the figure, $OBE$ is connected on other $OBE$s and $ITS$ devices, $TMC$, and most importantly $RSE$. Suppose all the nodes $N$=$\{$OBE,SA,...,DCC,LC$\}$ (see Table~\ref{tab_metrics2}). Given, 
\begin{eqnarray}
p(OBE)=\sum_{N/{OBE}}{\prod_{N/OBE}{p(OBE|pa(OBE))}} \\
p(OBE_{10})=\prod_{i\in I}{p(OBE_i)\prod_{i=1}^{10}{p(OBE_{i}|pa(OBE_{i}),OBE_{i-1})}}
\label{eq_pmfc}
\end{eqnarray}
where, $pa(OBE)$ and $pa(OBE_i)$ represent the parental variables of on-board equipment in single vehicle and CACC setting, respectively. Simply, when nodes are incorporated, Eq.~(\ref{eq_pmfc}) follows, 

\begin{eqnarray}
p(OBE)=\sum_{(RDS,RDD,...,BSM)}{p(OBE,RDS,RDD,...,BSM)} \nonumber \\ 
=p(OBE|RDS,RDD,...,VCA)p(VCA|VVV,...,HVS)...p(BSM) 
\label{eq_pmfobe}
\end{eqnarray}

Probabilities in Eq.~(\ref{eq_pmfobe}) from BNs are propagated (posteriors)as before. Example of propagated conditional distributions are given in Table~\ref{tab_obe}. Marginal probabilities of states of $OBE$ are calculated. When all input propagated, \textit{none} and \textit{low} higher than impacting $OBE$ states, hence, CACC stability. The impact on the $10^{th}$ $OBE$ is also propagated reasonably. The distributions are expected as attacks reaching to the critical node after detection layers is low. Overall, framework is promising if a detailed attack model is incorporated, the risk will be reflected realistically.  
\begin{table}[h!]
\centering
\caption{Propagated $P(OBE)$ and $P(OBE_{10})$}
\label{tab_obe}
      \centering
\scalebox{0.95}{
\begin{tabular}{rrrrrrr}
  \noalign{\smallskip}\hline\noalign{\smallskip}
 $OBE$ & 0.946 & 0.0538 & 0.0000215 & 0.00000332 & 0.00000299\\ 
  $OBE_{10}$ & 0.942 & 0.0582 & 0.0000673 & 0.0000192 & 0.0000174\\ 
    \noalign{\smallskip}\hline\noalign{\smallskip}
\end{tabular}
}
\end{table}

Probabilities (posteriors) in Eq.~(\ref{eq_pmfsc}) from BNs are calculated by the junction tree algorithm with computational complexity of $O(n^3)$. The algorithm actually yields the multiplication of conditional probabilities including the independence, thus, a simplified version. This algorithm, on directed acyclic graphs, first marries parent nodes. It then triangulates and removes unnecessary links through generation of cliques, assignments, and message propagation (\cite{barber2012bayesian}). 

Note that parent nodes $p(DOSR)$ require probability assumptions for uncertain evidence setting. In this paper, these scales are adopted from (\cite{petit2015potential}) and assigned from (\cite{cvss}). Specifically, defined attack vector levels are $\{low,medium,high\}$ ($L,M,H$) and state probabilities for nodes such as $OBE$ are $\{none,low,medium,high,critical\}$ ($N,L,M,H,C$) with corresponding maximum probability values of $\{0.25,$ $0.50,0.75\}$ and $\{0.01,0.39,0.69,0.89,1.0\}$ respectively. In order to address the sensitivity of these probabilities are generated randomly from uniform distribution and resulting posteriors are obtained as probability distributions for each state rather than a point value. Moreover, resulting expected utility for vulnerability are calculated out of $[0,10]$ scale where up to $0.1$ shows no risk or impact to $10$ is the highest vulnerability. Example of propagated conditional distributions are given in Table~\ref{tab_obe} (\cite{hojsgaard2012graphical}).
\subsection{Intelligent Signal Application}
\begin{figure*}[h!]
\centering
\begin{subfigure}{.487\textwidth}
 \centering
\fbox{\includegraphics[width=0.99\linewidth]{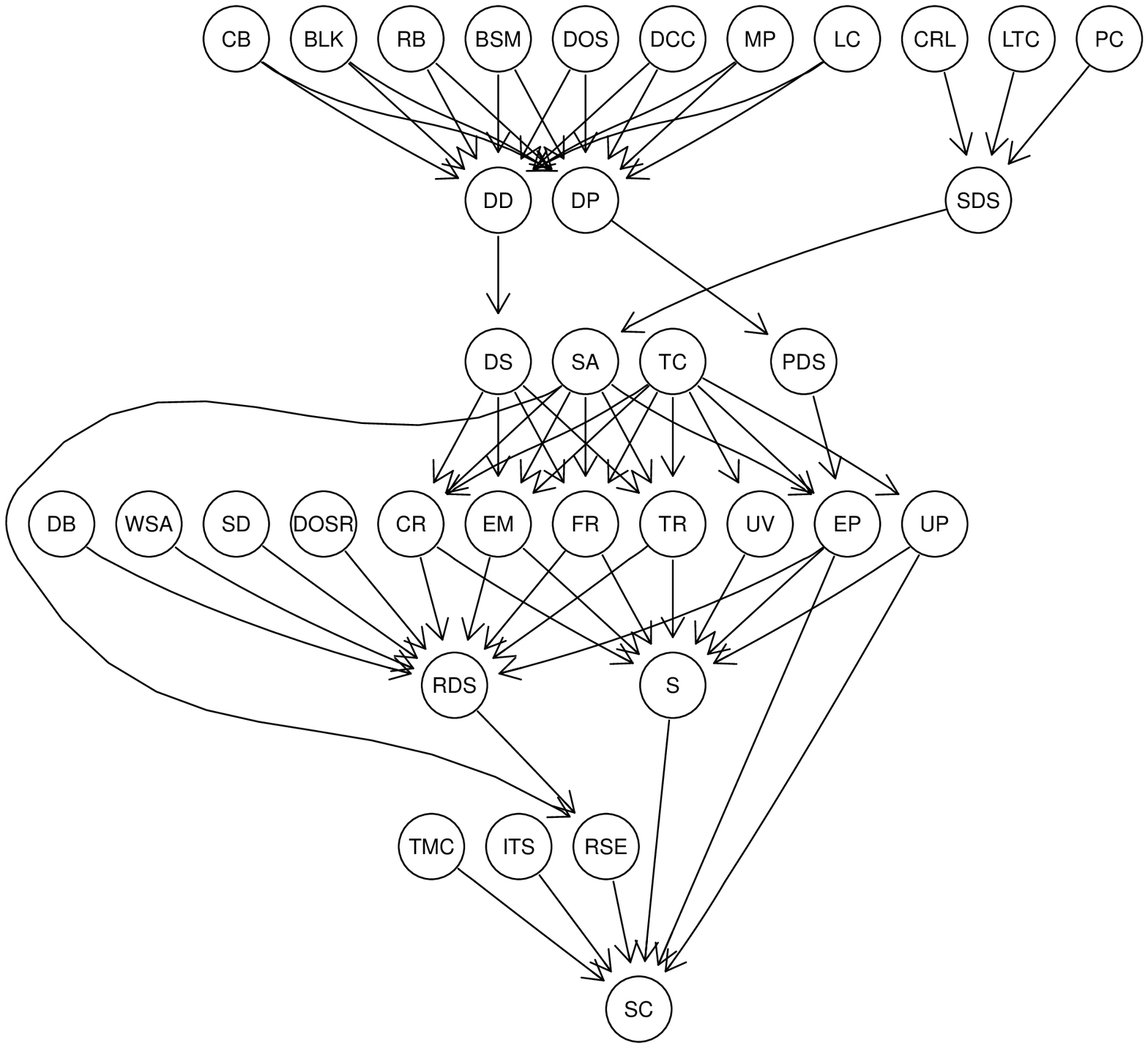}}
  \caption{w Sensor}
  \label{fig_bnsc}
\end{subfigure}
\begin{subfigure}{.47\textwidth}
\centering
  \fbox{\includegraphics[width=0.99\linewidth]{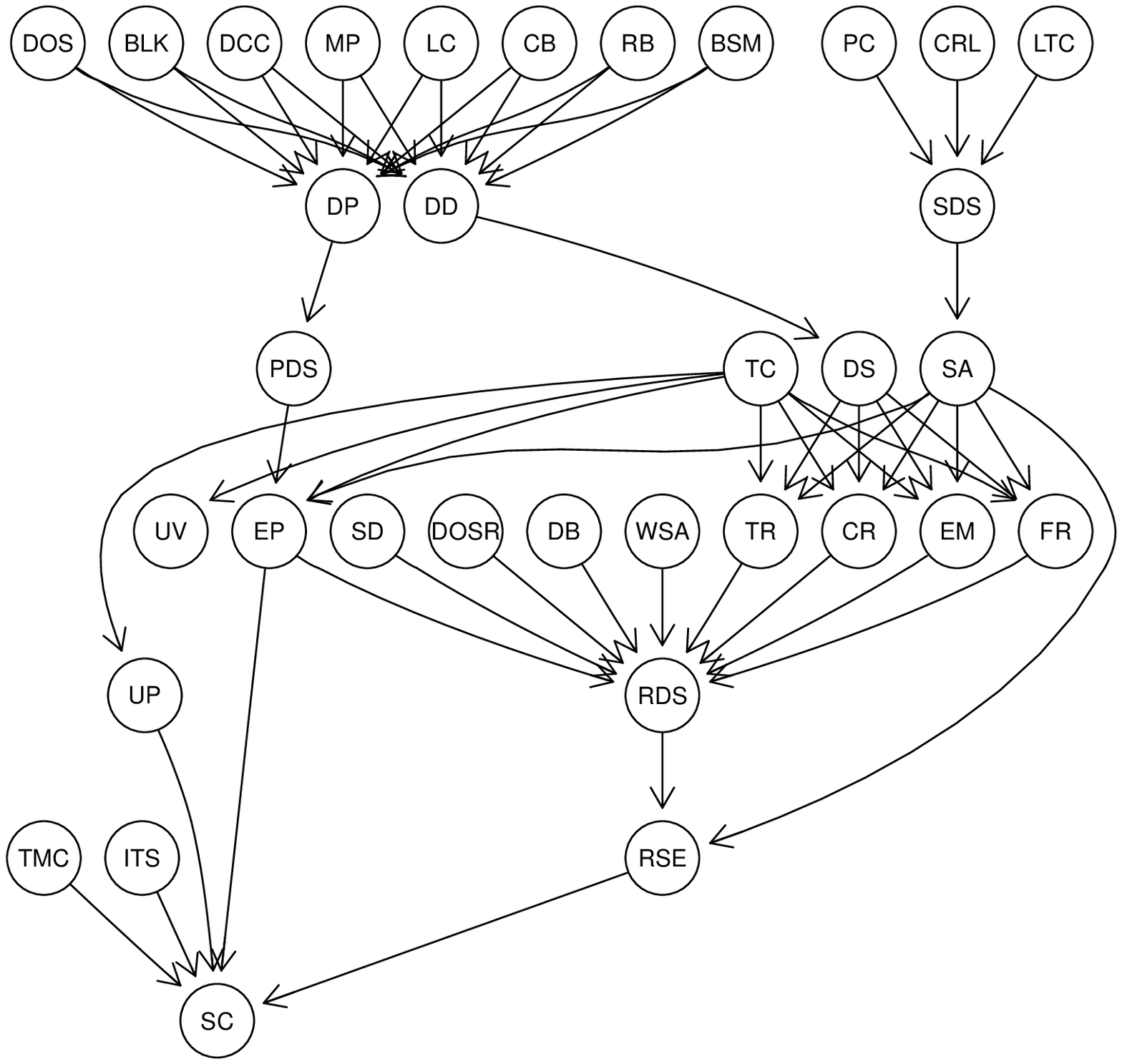}}
\caption{w/o Sensor}
  \label{fig_bnwos}
\end{subfigure}%
\caption{Bayesian Network with single signal controller}
\label{fig_bns}
\end{figure*}

In modeling, constraints of the system or assumed variables, their description, and states are given similarly in Table~\ref{tab_states} to define nodes on designed BNs. Details and metrics can be found in (\cite{comert2018modeling}). These attacks and metrics on simplified node set with assumed attack surfaces are listed in Table~\ref{tab_surf} are used to generate network in Fig.~\ref{fig_bns}. 
\begin{table}[h!]
\centering
\caption{Nodes on the designed ISIG BNs with assumed states}
\label{tab_states}
\scalebox{0.75}{
\begin{tabular}{lll}
  \hline
node & description & states \\ 
  \hline
 SC & signal controller & N,L,M,H,C \\  
 RSE & roadside equipment & L,M,H \\ 
 TC & traffic composition & UP,EP,UV,CR,EM,FR,TR \\ 
 UP & unequipped pedestrians & N,L,M,H,C \\ 
 EP & equipped pedestrians & N,L,M,H,C \\ 
 UV & unequipped vehicles & N,L,M,H,C \\ 
 CR & regular equipped vehicles & N,L,M,H,C \\ 
 EM & emergency vehicles & N,L,M,H,C \\ 
 FR & freight trucks & N,L,M,H,C \\ 
 TR & transit vehicles & N,L,M,H,C \\ 
 ITS & other ITS devices & L,M,H \\ 
 TMC & traffic management center & L,M,H \\ 
 S & sensor & True,False \\ 
 DD & detection by driver & L,M,H \\ 
 DP & detection by pedestrian & L,M,H \\ 
 DS & detection by system (vehs) & L,M,H \\ 
 PDS & detection by system (pedes) & L,M,H \\ 
 RDS & detection by system (RSE) & L,M,H \\ 
 SDS & detection by system (SA) & L,M,H \\  
   \hline
\end{tabular}
}
\end{table}

Fig.~\ref{fig_bnsc} shows the designed BN under assumed dependencies among the nodes similarly given in Table~\ref{tab_metrics2} (see\cite{comert2018modeling}). This network is a BN with single signal controller including a sensor (additional detection) node. This is critical as reliability of $SC$ could increase via additional ground truth check as well as control via this sensor which is quantified using simulation in section \ref{sctsims}. However, it requires additional cost. Similarly, detection nodes can be seen as cost for monitoring and delay of communications. In the figure, $SC$ is connected on pedestrian push buttons (assumed no impact on $SC$ state), other $ITS$ which can be other signals, $TMC$, $S$, and most importantly $RSE$. In the BN, modes of regular passenger vehicles, emergency vehicles, freight, transit, motorcycles, and other motorized modes are also shown. Now, suppose all the nodes can be shown as a set of $N$=$\{SC,SA,...,DCC,LC\}$. Given, 

\begin{equation}
p(SC)=\sum_{N/{SC}}{\prod_{N/SC}{p(SC|pa(SC))}}
\label{eq_pmf}
\end{equation}
where, $pa(SC)$ represents the parental variables of signal controller. Simply when nodes are incorporated, Eq.~(\ref{eq_pmf}) follows, 
\begin{eqnarray}
p(SC)=\sum_{(S,RSE,...,WSA)}{p(SC,S,RSE,...,WSA)} \nonumber \\ 
=p(SC|S,RSE,...,WSA)p(RSE|RDS,...)p(S|UV,UP,EP,...,TR)...p(WSA) 
\label{eq_pmfsc}
\end{eqnarray}

As an example joint distribution involving $SA$ risks can be simplified after conditional independence and written as in Eq.~(\ref{eq_pmfsa}). Inference on $p(SA)$ can be deduced after summing up for $(SDS,PC,CRL,LTC)$.
\begin{eqnarray}
p(SA,SDS,PC,CRL,LTC)=p(PC)p(CRL)p(LTC)p(SDS|PC,CRL,LTC)p(SA|SDS)\nonumber \\ 
p(SA)=\sum_{(SDS,PC,CRL,LTC)}{p(PC)p(CRL)p(LTC)p(SDS|PC,CRL,LTC)p(SA|SDS)}
\label{eq_pmfsa}
\end{eqnarray}

Example of propagated conditional distributions are given in Table~\ref{tab_scs}. In the table, for different state of $RSE$, probabilities of states of $SC$ are calculated. Notice that they are all high for \textit{none} as state of other nodes are also impacting $SC$ state. If $S$ is $false$, $p(SC=none)$ decreases. Similar probabilities can be calculated also for other higher nodes. For instance, detection for $RSE$ can be calculated as $P(RSE|RDS)$. Reasonably from the table, if detection state is $low$, then $RSE$ can be impacted high else impact on $RSE$ is $low$.   

\begin{table}[h!]
\centering
\caption{Propagated $P(SC|RSE)$, $P(SC|S)$, and $P(RSE|RDS)$ with and without sensor $p$=$100\%$}
\label{tab_scs}
\scalebox{0.75}{
\begin{tabular}{rrrrrrrrrrr}
  \hline
 & \multicolumn{6}{c}{SC} & \multicolumn{4}{c}{RSE} \\
& & none & low & med & high & cri& & low & med & high \\ 
  \hline
 RSE & low & 0.929 & 0.0698 & 0.0003 & 0.0004 & 0.0002& RDS& 0.468 & 0.162 & 0.370 \\ 
  & med & 0.900 & 0.0669 & 0.0176 & 0.0140 & 0.0011 & & 0.915 & 0.045 & 0.040 \\ 
  & high & 0.897 & 0.066 & 0.0013 & 0.0206 & 0.0143 & & 0.994 & 0.003 & 0.003 \\ 
   \hline
   S& true & 0.929 & 0.069 & 0.0003& 0.0004 & 0.0003 &&&&\\ 
    & false & 0.729 & 0.266 & 0.0012 & 0.0016 & 0.0011 &&&&\\
    \hline 
RSE & low & 0.735 & 0.261 & 0.0012 & 0.0015 & 0.0010& RDS& 0.444 & 0.163 & 0.393 \\ 
    & med & 0.671 & 0.238 & 0.0468 & 0.0394 & 0.0042 & & 0.913 & 0.046 & 0.041 \\ 
    & high & 0.662 & 0.235 & 0.0047 & 0.0593 & 0.0385 & & 0.995 & 0.003 & 0.002 \\ 
       \hline
\end{tabular}
}
\end{table}
\section{Simulations}
\label{sctsims}
Three main type of simulations are run to evaluate modeling and quantify risks. Monte Carlo simulations (MCSs) are used to quantify variations on Bayesian Networks by running many cycles and replications. Microscopic (realistic including vehicle following) and horizontal (simpler and less realistic) traffic simulations are run to quantify impact of attacks in terms of common performance measures. MCSs are normally alternative to analytical calculation methods such as BNs. With MCSs, time aspect can be handled better.
\subsection{Monte Carlo Simulations}
In order to further analyze the cyber-physical model, impact on a CACC application, the BN in Figs.~\ref{fig_bn1} and \ref{fig_bn10} is propagated through speed changes of leading vehicles in the example network. In terms of analyzing the impact on intelligent signal applications, signal controller is incorporated via service times at the signalized networks of number of signals in the example network, as shown in Figs.~\ref{fig_bns} and \ref{fig_netpt}. Primary objectives are quantification and identification of possible problems and mitigation efforts for future research. These yield risk areas, impacts, and possible solutions for more resilient systems. MCSs are given using BNs including an isolated intersection and single OBE as well as replications of complete BNs mimicking a network (i.e., pattern). MCSs generate probability distributions and expected utility as well as its dispersion so that confidence intervals can be generated. Pattern approach generates a Bayesian framework for posterior probability of a critical node common among these signals such as security (certificate) authority considering it as a parameter. 
\subsubsection{CACC}
\begin{figure}[h!]
\centering
\includegraphics[width=0.8\linewidth]{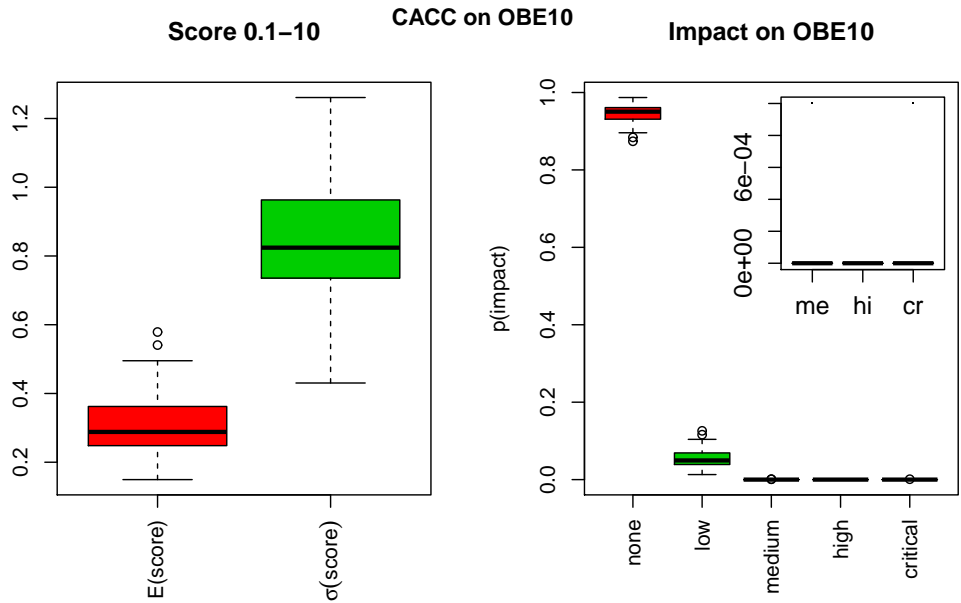}
\caption{Simulated score values from BN of Fig.~\ref{fig_bn10} for $10^{th}$ OBE}
\label{fig_pmss}       
\end{figure}

In Fig. \ref{fig_pmss}, probabilities, expected value, and standard deviation of vulnerability of OBE are presented. Monte Carlo simulations are run for $10,000$ cycles. Boxplots are used in order to show the impact distribution for $50$ replications. Metrics are assigned randomly from uniform distribution. Probabilities that are used to calculate the moments. Probability of OBE with no effect state is about $94\%$ and about $5\%$ low impact. These values are changing within $[0.90,0.99]$ and $[0.00,0.15]$, respectively changing the central tendency and the dispersion of vulnerabilities. Expected value certainly falls mostly into low impact $>0.10$ region. Median of standard deviations is about $0.95$. Thus, a simple confidence interval for mean level of $0.30$ with mean standard deviation of $0.80$ from $\mu\pm2\sigma$ would fall within $[0.00,1.90]$. This is still in the low impact scale. In order to check the effect of randomly generating all metric values. These values are only assigned for $50$ replications but not for each node on BNs. This approach is exploring enough of the metric ranges and resulting higher dispersion. Hence, all random approach would represent more robust results.
\subsubsection{Intelligent Signals}
Probabilities, expected value, and standard deviation of vulnerability of $SC$ are presented. Boxplots are used in order to show the ranges within $200$ replications. Probabilities that are then used to calculate the moments. Probability of a node (e.g., $SC$) being in no effect/$none$ level is about $95\%$ and about $5\%$ $low$ impact when a redundant sensor is included. These values are changing between $[0.80,0.99]$ and $[0,0.20]$ respectively changing the central tendency and the dispersion of vulnerabilities. Expected value certainly falls mostly into $low$ impact $>0.1$ region. Median of standard deviations is about $0.95$. Thus, a simple confidence interval for mean level $0.37$ with mean standard deviation $0.93$ $\mu\pm2\sigma$ would be within $[0,2.23]$ which is still in the $low$ impact scale.  
\begin{figure}[h!]
\centering
\begin{subfigure}{0.8\textwidth}
 \centering
\includegraphics[width=0.9\linewidth]{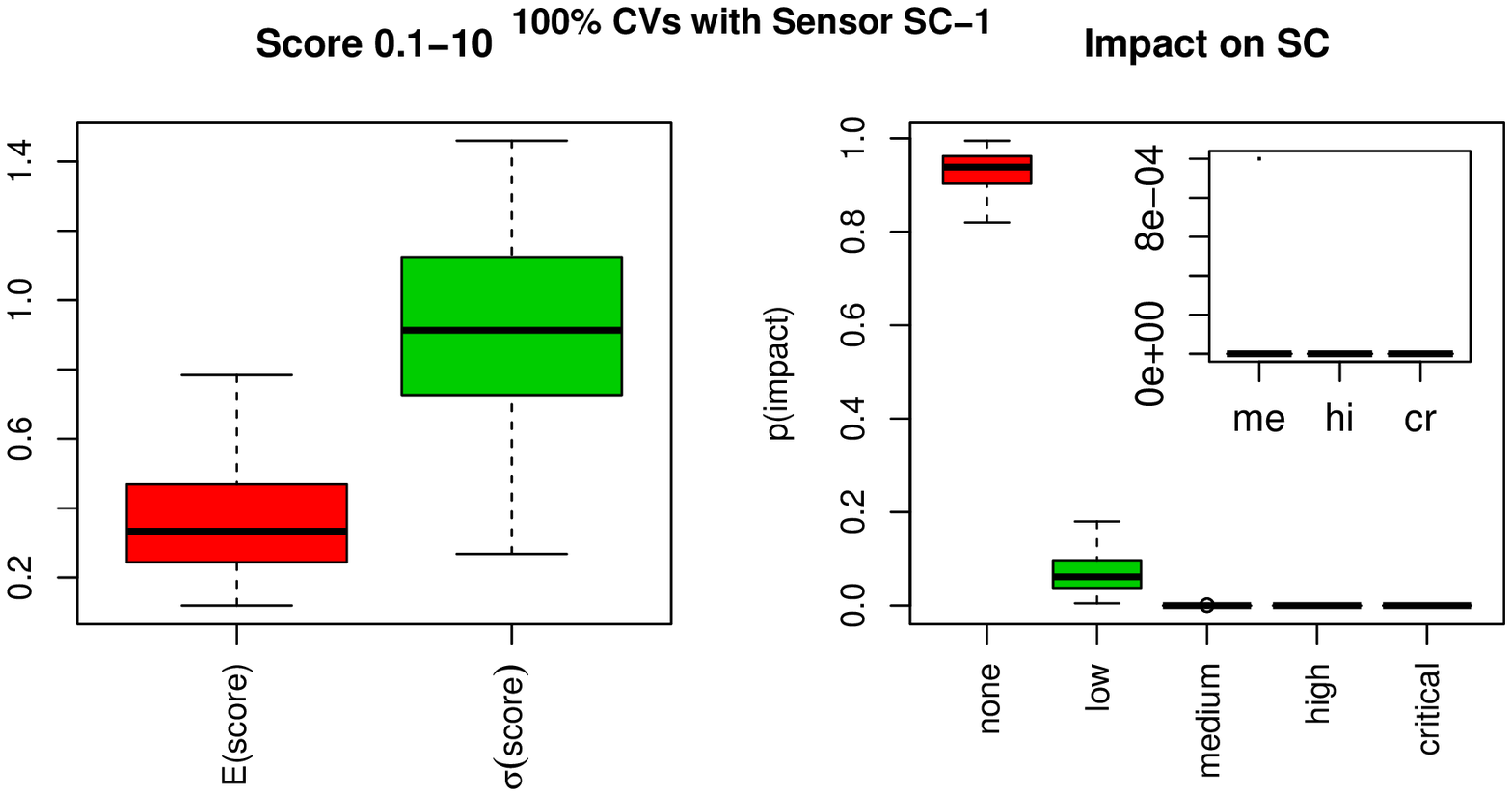}
  \label{fig_ptsc1}
\end{subfigure}
\centering
\begin{subfigure}{.8\textwidth}
 \centering
\includegraphics[width=0.9\linewidth]{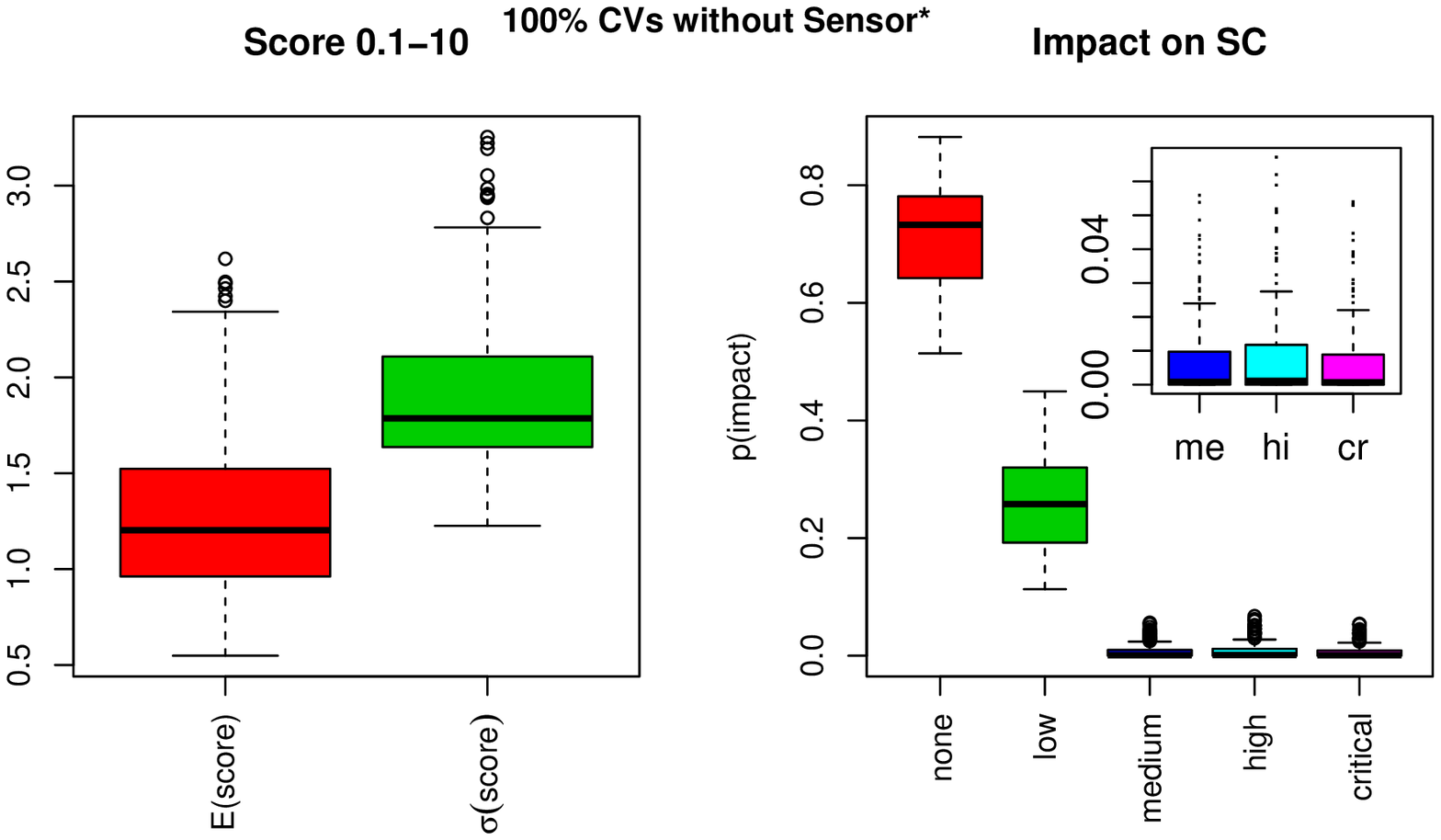}
  \label{fig_wosr}
\end{subfigure}
\caption{Simulated score values from BN with and without sensors}
\label{fig_wos}
\end{figure}

Similar values are given in Fig.~\ref{fig_wos} for without sensor case. Results reveal much higher risk levels and variations. Risk likelihood is still mainly within $low$ level, however, it is more probable with $low$ state than $none$. Confidence interval in this case would be $[0.00,4.81]$ with mean and standard deviation values $1.19$ and $1.81$ respectively. In this case, vulnerability falls into much serious $medium$ level. Same as before less random version yields higher dispersion with confidence interval $[0.00,5.03]$ with means $1.28$ and $1.87$.  
\begin{figure}[h!]
\centering
\fbox{\includegraphics[scale=.7]{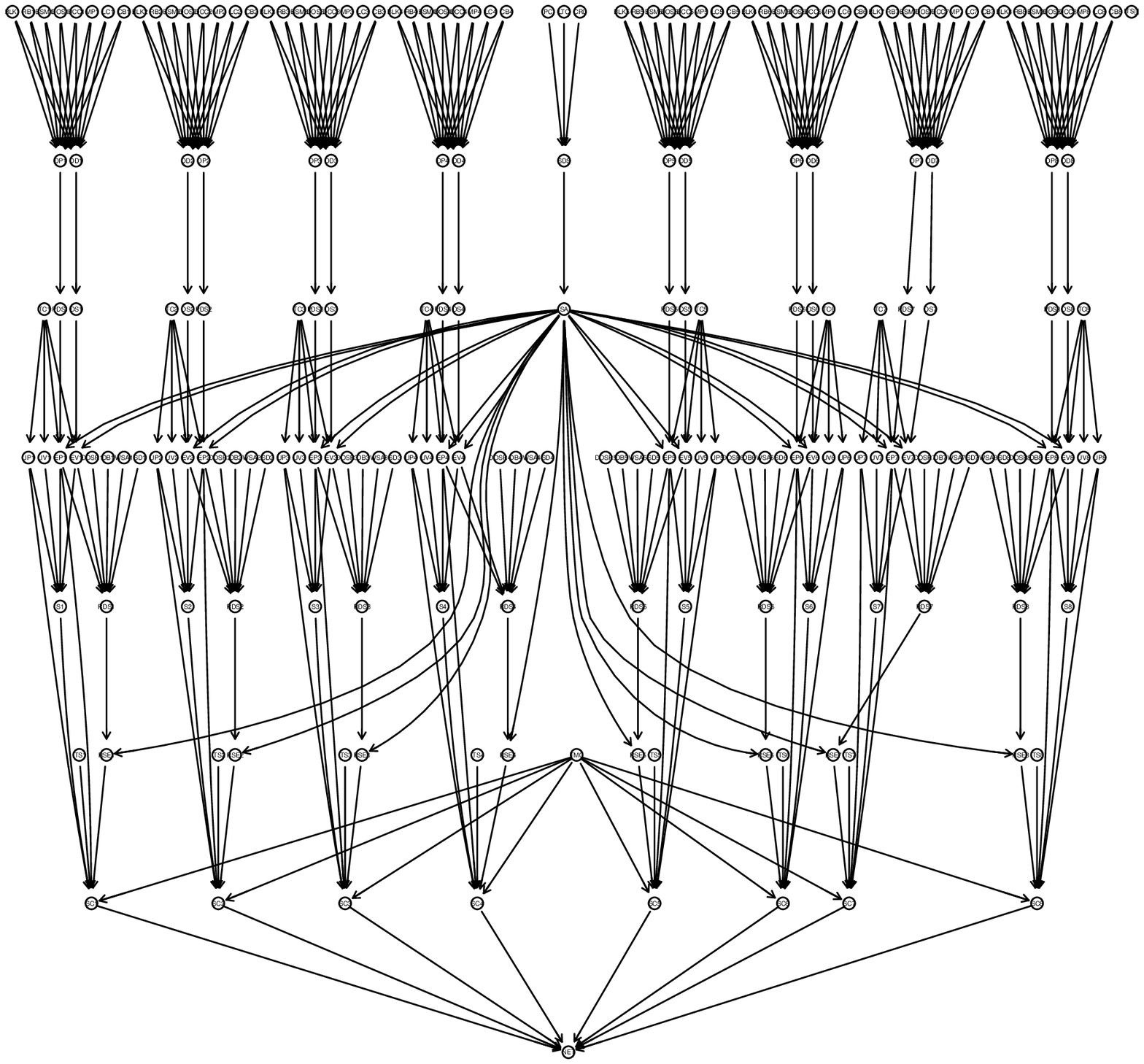}}
\caption{BN with 8 signal controllers}
\label{fig_netpt}       
\end{figure}

\begin{figure}[h!]
\centering
\begin{subfigure}{0.8\textwidth}
\centering
  \includegraphics[width=0.9\linewidth]{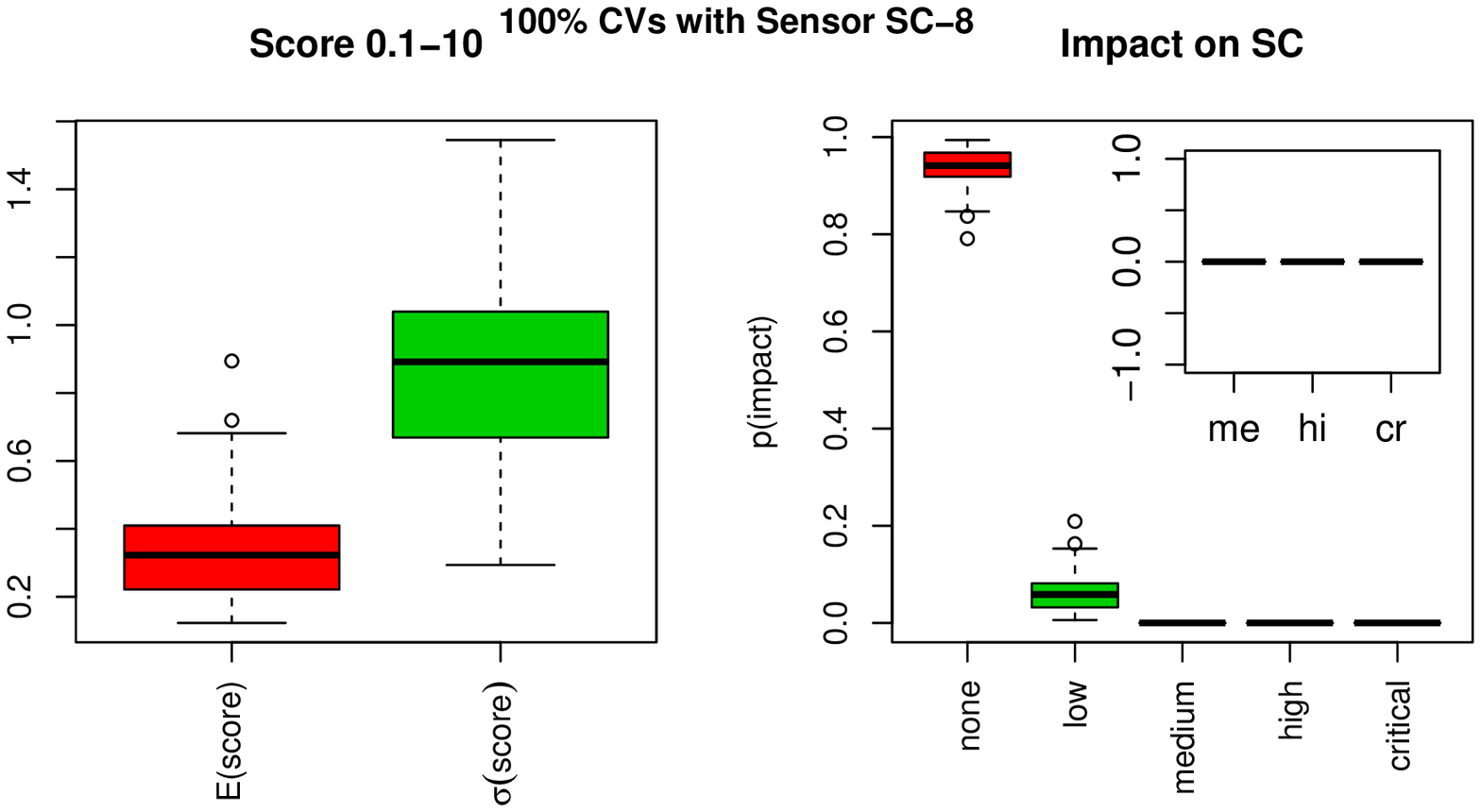}
\caption{Eighth $SC$}
  \label{fig_ptsc8}
\end{subfigure}\\
\begin{subfigure}{0.8\textwidth}
\centering
  \includegraphics[width=0.9\linewidth]{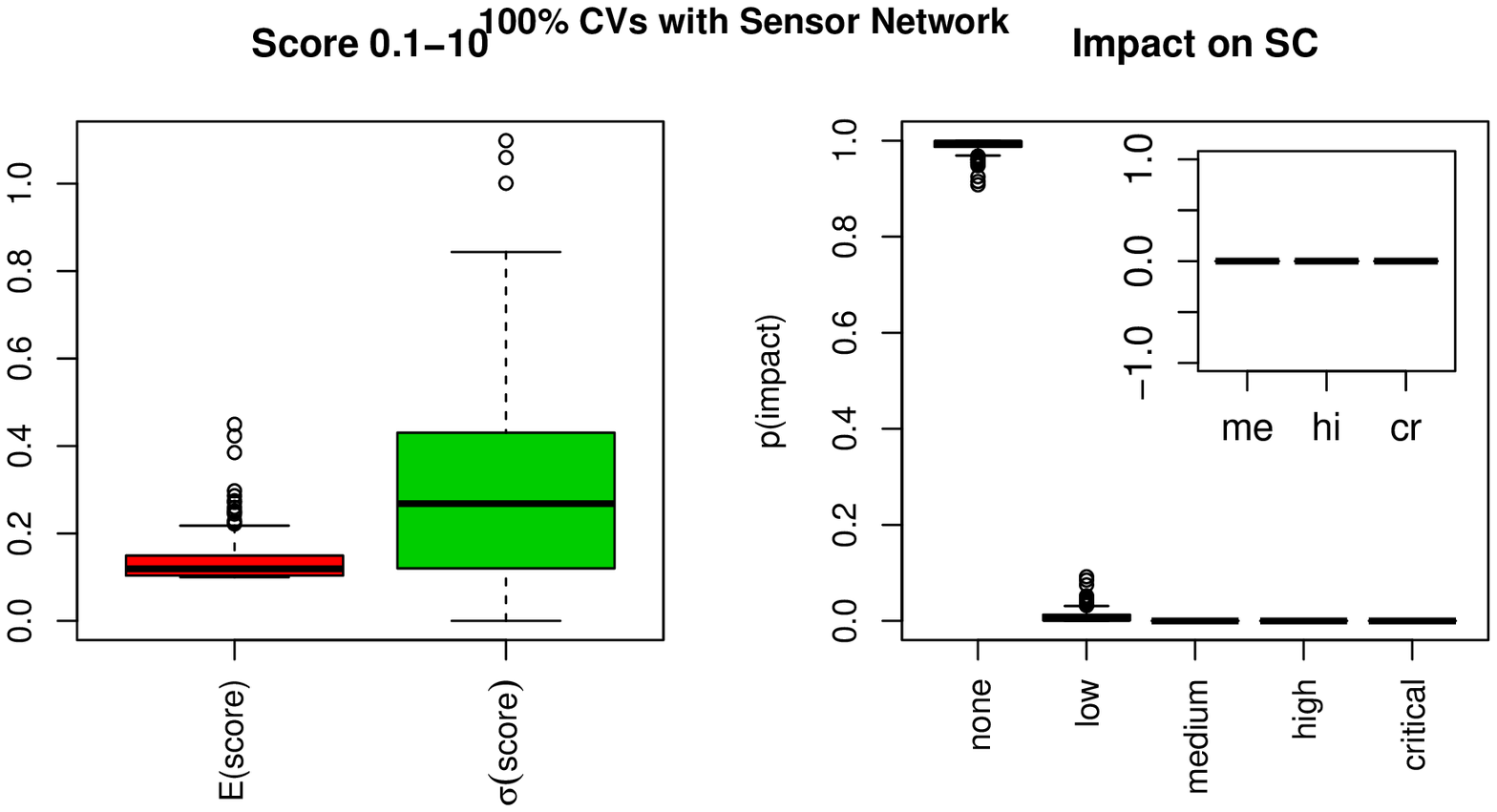}
\caption{Entire network}
  \label{fig_ptnet}
\end{subfigure}%
\caption{Simulated Risk values from BN with sensors}
\label{fig_pattern}
\end{figure}

Fig.~\ref{fig_netpt} shows pattern of $8$ BNs in Fig.~\ref{fig_bns}. Essentially, it is a network of controllers with common security certificate authority ($SA$), traffic management center, and other $ITS$ devices. Vulnerability of individual signal controllers (Fig.~\ref{fig_ptsc8}) as well as entire network (Fig.~\ref{fig_ptnet}) is presented in Fig.~\ref{fig_pattern}. From the figure, similar risks for individual signal controllers can be seen. However, risk reduces in a network setting. In this system, dependency is represented only through $SA$, $TMC$, and $ITS$ nodes. Thus, overall risk is much less than a single node. It is also observed that a risk in a signal controller increases in a serial framework based on defined transitions among signal controllers which is intuitive.
\subsection{CACC Microscopic Simulation}
\label{scsctts}
Microscopic traffic simulation environment, VISSIM, generates vehicles with exponential interarrival times at the origin that traverse links based on realistic vehicle following behavior according. Headways change as vehicles move along the network based on vehicle composition, vehicle characteristics, driving behavior, number of lanes, and other network settings. Similar to the complex real-life traffic systems with multiple parameters to control, detailed analysis could be done in order to make sure a fully accurate comparison under different scenarios which is beyond the scope of this paper.  
\begin{figure}[h!]
\centering
\fbox{\includegraphics[width=0.75\linewidth]{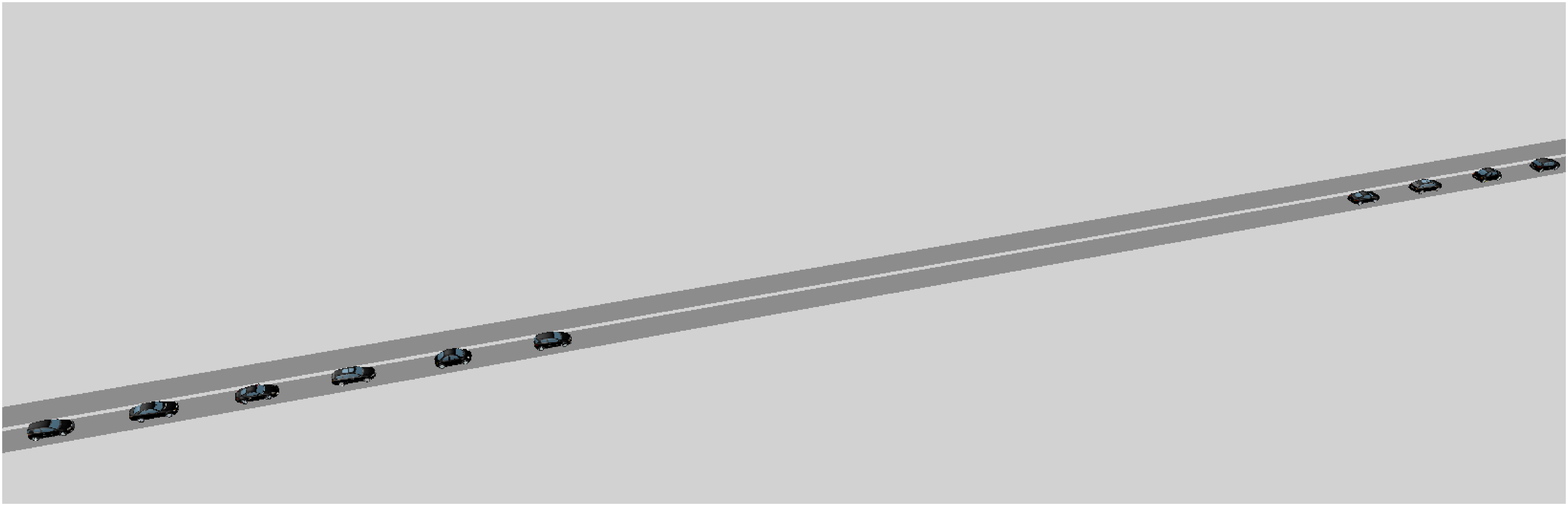}}
\caption{CACC VISSIM Simulation}
\label{fig_vissim}       
\end{figure}

In this study, VISSIM simulations were controlled using COM interface and Matlab integration. For a basic evaluation purpose, default parameters for vehicle characteristics (e.g., acceleration/deceleration, type $100$ car). Impact of the state probabilities on a $1.85$ kilometers ($km$) section of a two-lane highway is shown in Fig.~\ref{fig_vissim} using a microscopic simulation VISSIM with $4$ platoons of $10$ vehicles ($v$) equipped with OBE. On average $65$ $s$ to traverse the link. Vehicles at source are generated from $N(10, 1)$ $m$ apart (approximately $0.5$ $s$ following time headways) and platoons generated 50 meters ($m$) apart from each other. Vehicles do accelerate and decelerate based on platoon leader’s speed information passed at $0.1$ $s$ intervals, without attack, all platoons stay in tack and stable. Simulation run lengths are $1$ hour ($h$) where new platoons are generated as other platoons are discharged from the network. 

For simplification, the facility is not interrupted and communications are not included. Hence, changes are incorporated only on speed passed within basic safety messages. Average delay and number of stops are demonstrated for no attack and attack scenarios. As the objective is reflecting the impact of disturbance, possible interruptions are generated from the probability distributions BNs. Impacts are reflected on speed on kilometer per hour ($kmh$) with $(none,low,med,high,cri)$ corresponds to $(0.0,N(5,5),N(80,120),N(160,240),N(320,480))$ $kmh$ with average probabilities OBEs $(0.9460, 0.0538$, $2.15\times10^{-5}, 3.0\times10^{-6}, 3.0\times10^{-6})$. 
\subsection{Simulations for Intelligent Signals}
\label{sctts}
\begin{figure}[h!]
\centering
\includegraphics[scale=.75]{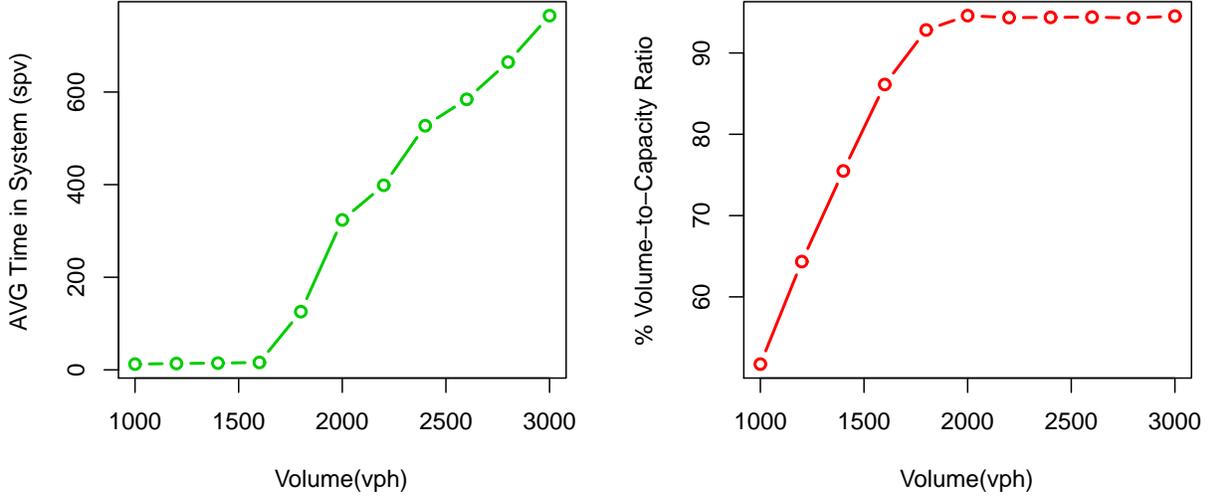}
\caption{Simulation results of isolated intersection}
\label{fig_simint}       
\end{figure}

Impact of the state probabilities on a network of signals is shown using ProModel Process Simulator with $25$ controllers. For simplification, signals are assumed to be fixed and communications are not included. Hence, changes are incorporated only through service times. Average queue lengths (in number of vehicles ($v$)) and waiting times (in seconds ($s$)) are demonstrated for regular and with impact scenarios. For set up simulations, first saturation level is determined. Comparing with the results from \cite{comert2016queue}, isolated intersection with two one way approaches is designed with $45$ seconds ($s$) green each approach and volumes=$\{500,600,700,...,2000\}$ vehicles per hour ($vph$) per approach. Similar results are observed with saturation at about $1000$ $vph$ per approach. Fig.~\ref{fig_simint} shows volume versus average and maximum queue lengths, average delays (average waiting time in queue and service) in seconds per vehicle ($spv$), and volume-to-capacity ratios (utilization). From the figure, saturation point can be seen at close to $2000$ $vph$ with capacity of $23$ vehicles per cycle and $1.95$ $spv$ discharge rate.    

\begin{figure}[h!]
\centering
\fbox{\includegraphics[scale=.4]{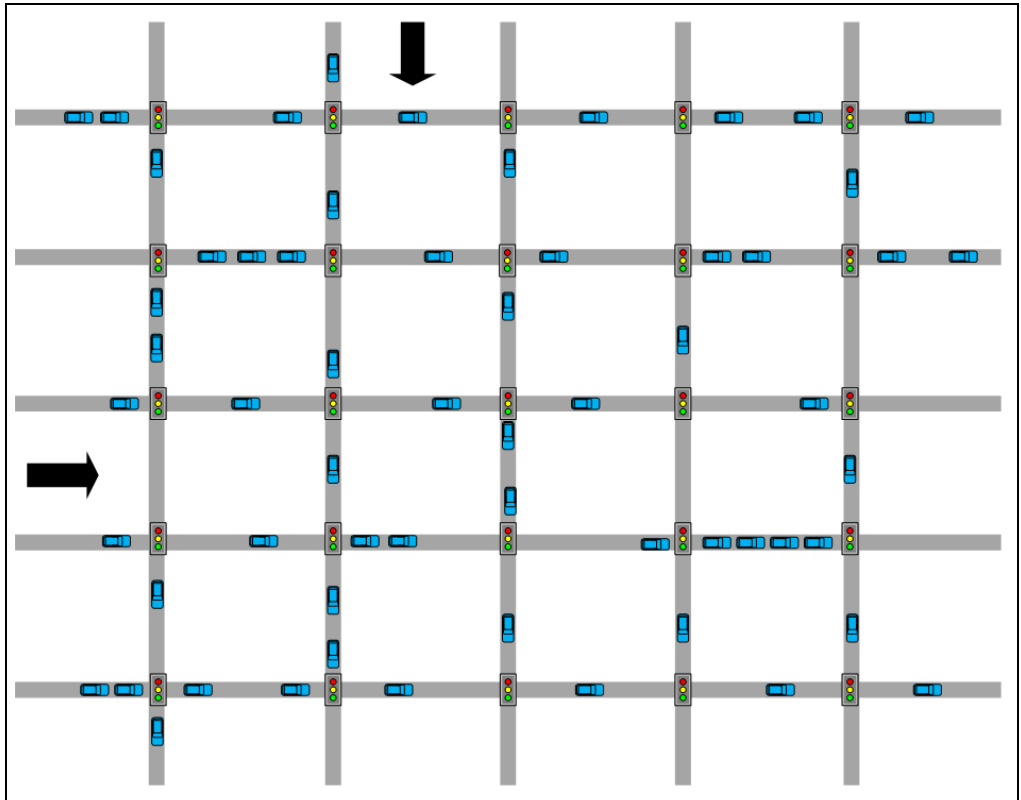}}
\caption{Network of signalized intersections}
\label{fig_net}       
\end{figure}

Next as shown in Fig.~\ref{fig_net}, a signalized network with $25$ intersections is designed. Each intersection is placed within $0.61$ $km$ taking exponentially on average $46$ $s$ to travel from one intersection to another. As the objective is reflecting the impact of disturbance, these dimensions represents a network of intersections with approach capacities that can accommodate maximum of $130$ average vehicle length per approach. Using the same volume rates up to $2200$ $vph$ per intersection, possible interruptions are generated from the probability distributions BNs. Impacts are reflected on $1.95$ $spv$ with $\{none,low,med,high,cri\}$=$\{$1.95,2.60,3.90,9.75,19.5$\}$ $spv$ with average probabilities with sensors results $\{$0.93,0.0685,0.00051,0.0005,0.00049$\}$ and without sensors $\{$0.727,0.261,0.00035,0.00046,0.00034$\}$, respectively. 
\section{Results and Discussions}
\label{sctne}
In this section, numerical results are presented for BNs with and without redundant surveillance technologies and with and without detectors. Monte Carlo simulations are run for $10,000$ cycles which results as state probabilities, expected and standard deviation of vulnerability. Simulations are repeated $200$ times to obtain distributions of these measures for connected vehicle market penetrations of $p=[0.01,1.00]$ which is constituted by $\{CR=0.80,EM=0.03,FR=0.07,TR=0.10\}$. Vehicle and pedestrian traffic are $95\%$ and $5\%$.   

\begin{figure}[h!]
\centering
\includegraphics[scale=.45]{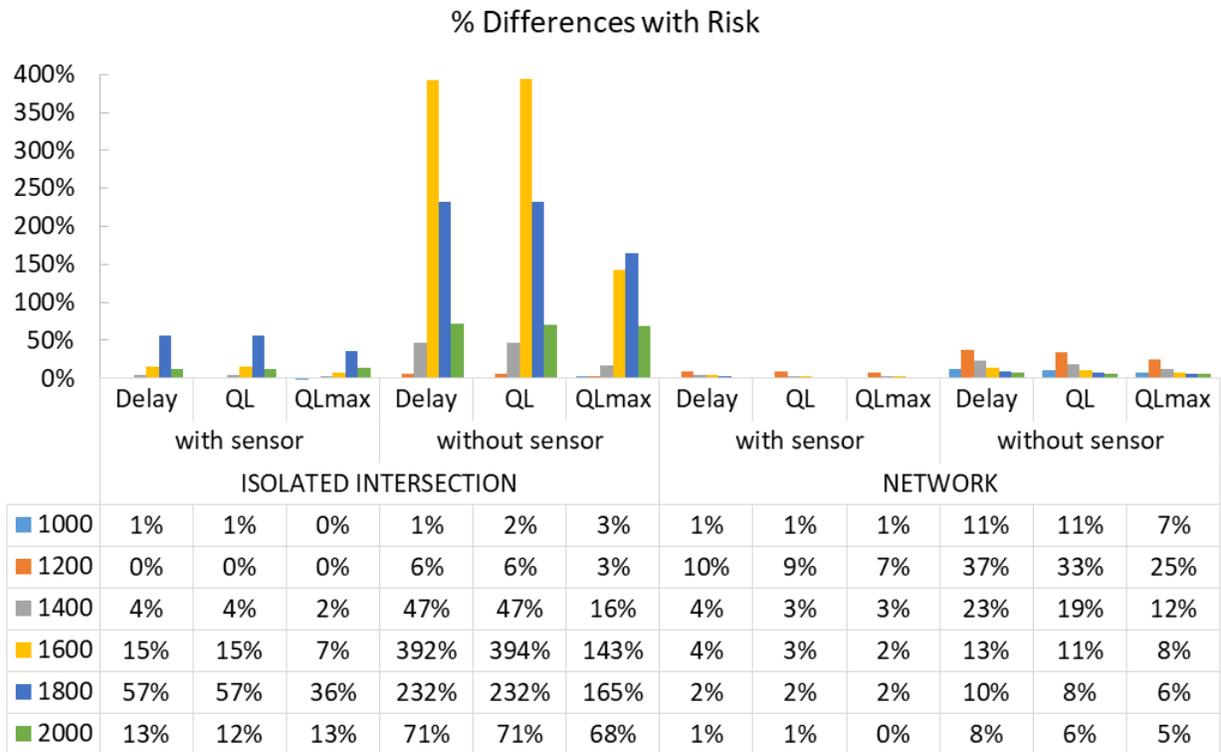}
\caption{Intersection performance measure differences when risks are incorporated}
\label{fig_pms}       
\end{figure}

Fig.~\ref{fig_pms} demonstrates overall summary of the impact of risk levels on queue lengths and delays obtained from process traffic simulations which can be viewed as realistic between microscopic and point queue simulations. Isolated results are adopted from (\cite{comert2018modeling}) and presented for comparison in this paper. Impact on single intersection is much higher about $15\%$ average QL, $10\%$ max queue length, and $15\%$ delay increase with a sensor. Without any sensor, changes are much higher with average $125\%$ average QL, $66\%$ max queue length, and $125\%$ delay across all volume levels. For network, the effect observed to be less average over all 25 signals, approaches, and volumes $3\%$ average QL, $3\%$ max queue length, and $5\%$ delay with sensor and $15\%$ average QL, $10\%$ max queue length, and $17\%$ delay without sensor case.

\begin{table}[h!]
\centering
\caption{Average differences of performance measures for ISIG}
\label{tab_summary}
         \scalebox{0.8}{
\begin{tabular}{lllllll}
\hline\noalign{\smallskip}
& \multicolumn{3}{c}{with sensor} & \multicolumn{3}{c}{without sensor} \\

   Type& Delay(spv)& Queue(v) & MaxQueue(v) & Delay(spv) & Queue(v) & MaxQueue(v) \\ 
\hline\noalign{\smallskip}
isolated & 9.82 & 2.52 & 4.21 & 82.28 & 21.03 & 28.78 \\ 
    & 15\% & 15\% & 10\% & 125\% & 125\% & 66\% \\ 

  network & 5.00 & 0.81 & 1.76 & 23.14 & 3.62 & 7.14 \\ 
    & 4\% & 3\% & 3\% & 17\% & 15\% & 10\% \\ 
\noalign{\smallskip}\hline\noalign{\smallskip}
\end{tabular}
}
\end{table}
Table~\ref{tab_summary} exhibits over all average differences and relative percentages. Average differences with sensor in queue lengths are $2.5$ and $0.8$ vehicles ($v$) for isolated and network respectively. In delays, they are $9.82$ and $5$ $spv$ and $4.21$ and $1.76$ $v$ in max queues.  

\begin{figure}[h!]
\centering
\includegraphics[scale=.85]{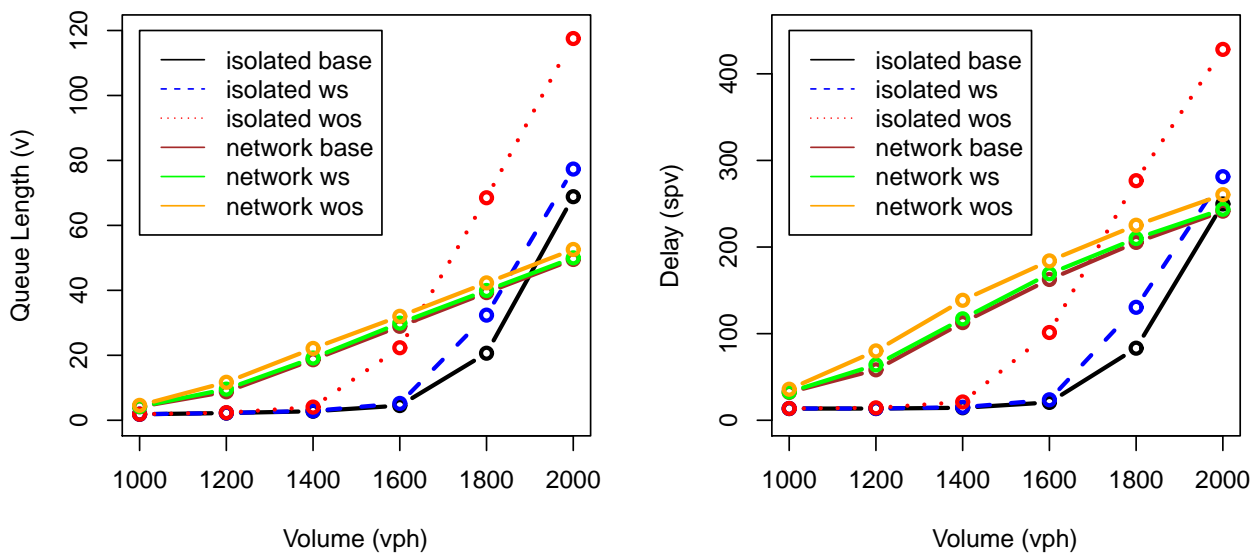}
\caption{Intersection performances for isolated (\cite{comert2018modeling}) and network cases}
\label{fig_qls}       
\end{figure}

Fig.~\ref{fig_qls} provides the overall evolution of performance measures for base, with sensor, and without sensor scenarios for isolated and network cases. Interestingly, up to $1600$ $vph$, network results are as expected and higher than isolated intersection values. After $1600$ $vph$, values for network simulations are getting less than that of isolated case. This observation could be due to lack of run length for network case when volume is higher meaning vehicles are not discharged as many at downstream signals compared upstream ones. Regardless, looking at the system performance after one hour with various volume levels, results reflect the impact of the incorporated risk which can be experienced at this level after one hour, but, certainly can vary in longer horizons.  

\begin{table}[h!]
       \centering
         \caption{Average differences of performance measures for CACC}
         \label{tab_summary2}
         \scalebox{0.8}{
         \begin{tabular}{llllll}
         
            State-Low& Avg Delay(s)& Avg Speed(kph) & Stops(v) & Total Delay(s) & \% Diff \\ 
         \hline\noalign{\smallskip}
         No Attack & 0.459 & 84.875 & 0 & 18.359 & - \\ 
          N(5,20)   & 0.664 & 84.816 & 0 & 26.552 & 45\% \\ 
         
           N(5,20) & 0.685 & 84.826 & 0 &  27.395 & 49\% \\ 
             N(5,5)& 0.722 & 84.830 & 0 &  28.877 & 57\% \\
             N(10,10)& 0.748 & 84.830 & 0  & 29.907 & 63\% \\
             N(40,60)& 8.300 & 80.307 & 13 &  331.986 & $\ge$100\% \\ 
         \noalign{\smallskip}\hline\noalign{\smallskip}
         \end{tabular}
         }
\end{table}

\begin{figure*}[ht]
\centering
\begin{subfigure}{.331\textwidth}
 \centering
\includegraphics[width=1.0\linewidth]{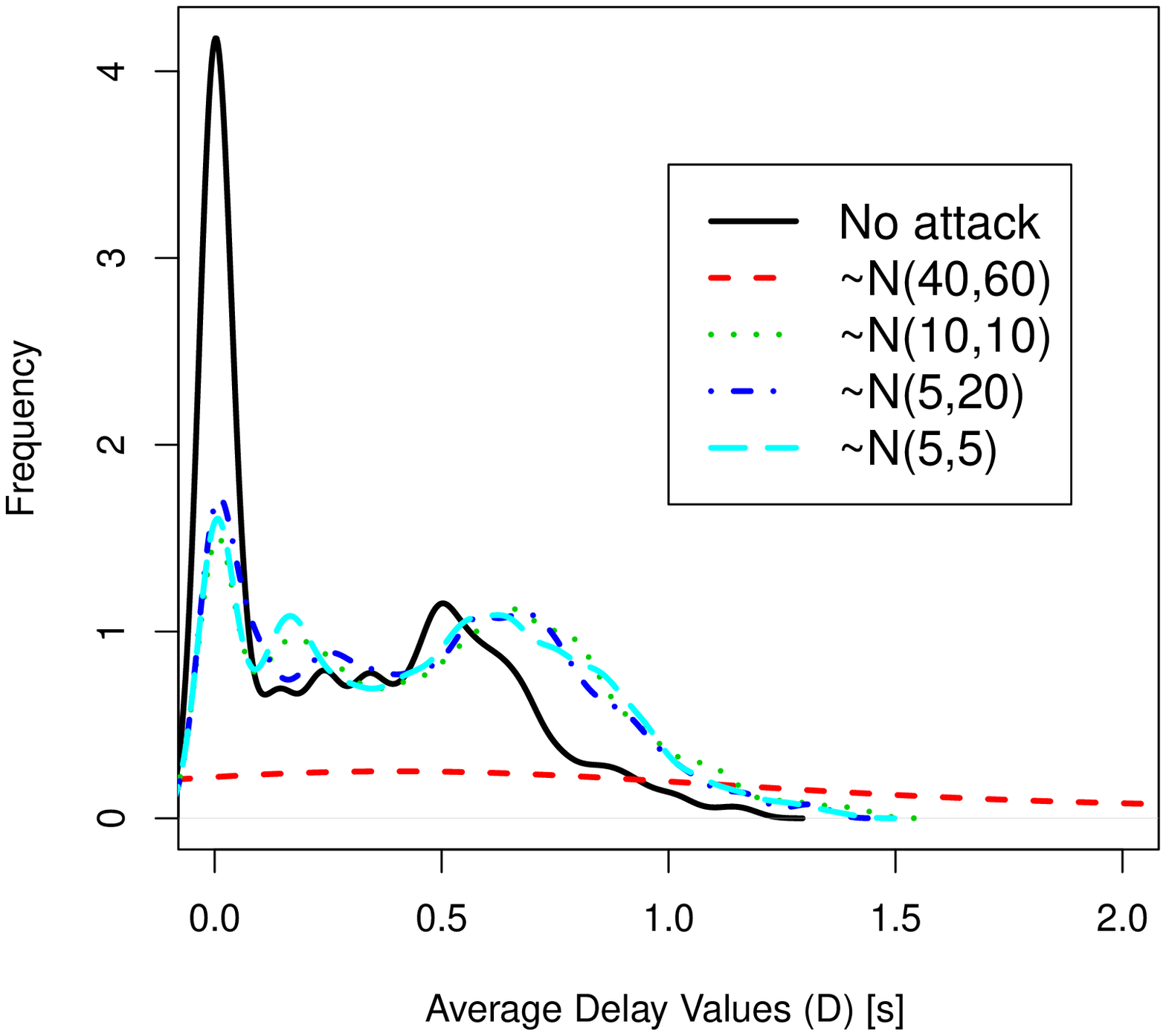}
  \caption{DOS attack}
  \label{fig_dd}
\end{subfigure}
\begin{subfigure}{.331\textwidth}
\centering
  \includegraphics[width=1.0\linewidth]{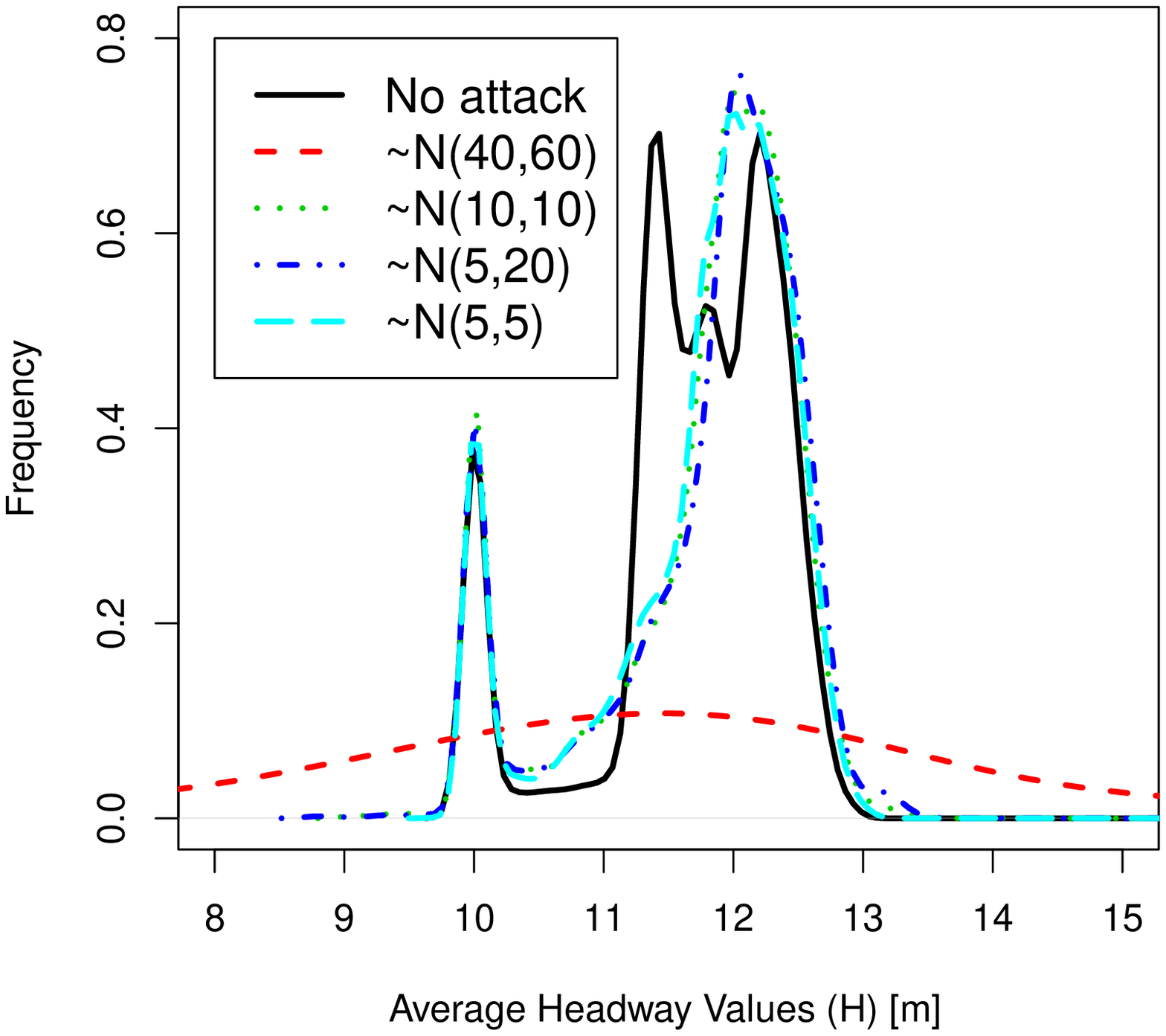}
\caption{Impersonation attack}
  \label{fig_dh}
\end{subfigure}%
\begin{subfigure}{.331\textwidth}
\centering
  \includegraphics[width=1.0\linewidth]{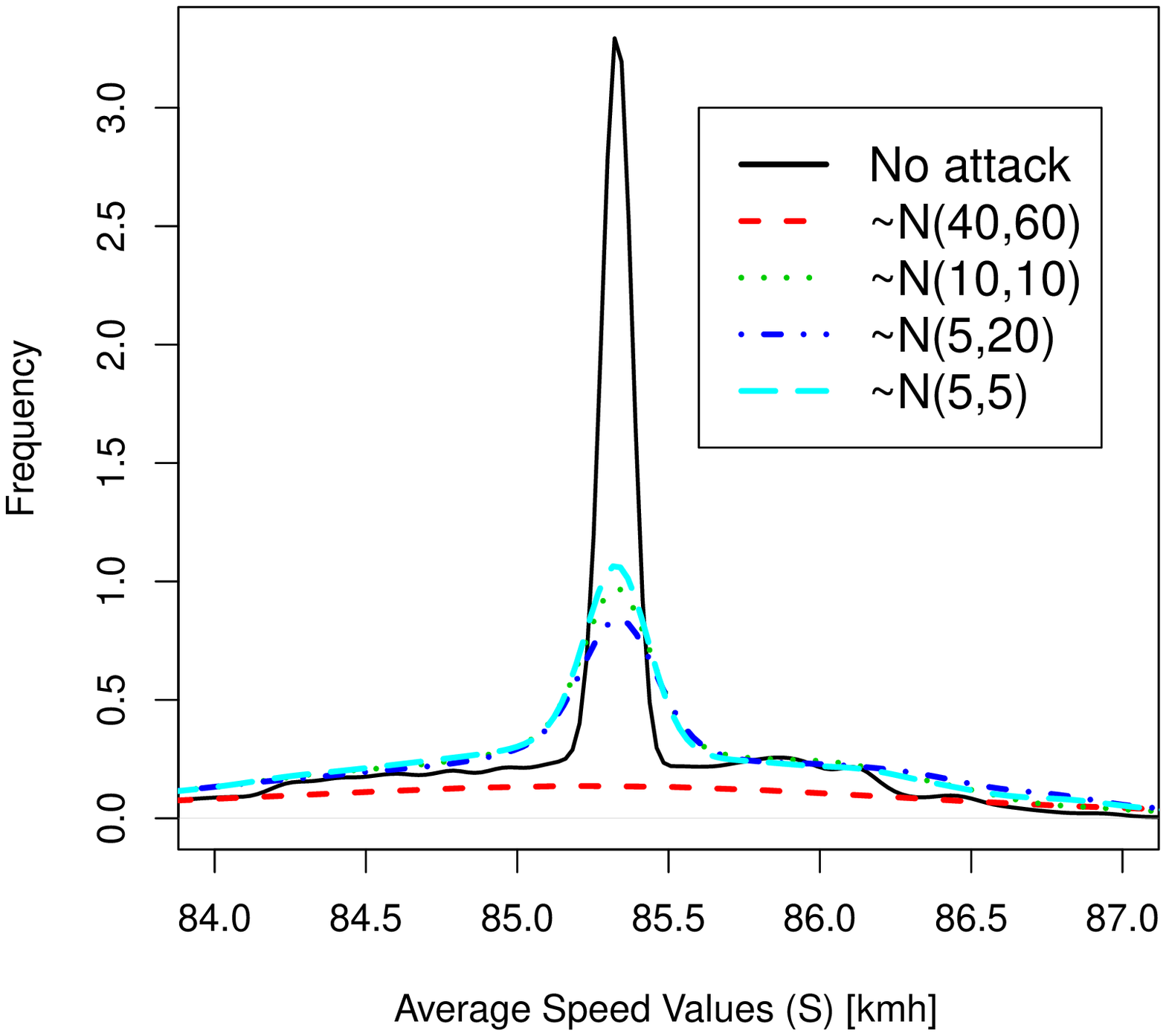}
\caption{False information attack}
  \label{fig_sd}
\end{subfigure}%
\caption{Impact of Attacks on CACC}
\label{fig_density}
\end{figure*}

Fig.~\ref{fig_pmss} demonstrates overall summary of impact of risk levels on delays obtained from process traffic simulations which can be interpreted as realistic between microscopic and point traffic simulations. Table~\ref{tab_summary2} exhibits over all average differences and relative percentages. Impact on platoons is much higher than $100\%$ average delay $s$ when low state is increased to $\sim N(40,60)$. When low state reduced to $\sim N(10,10)$ and $\sim N(5,5)$, impact decreases to approximately $5\%$. Note that probability of low impact on designed network is about $5 \%$. Any less than $95\%$ will impact the CACC system even if speed distribution is $\sim N(5,5)$. Impact of  different levels of attacks on average delay, headway ($ft$), and speed can also be seen from Figs.~\ref{fig_density} as weight over higher values of delay, larger variation of headways, and higher variation in average speed values.  
\section{Conclusions}
\label{sctconc}
In this paper, system level cyber attack risk and impact under connected vehicles framework using Bayesian Networks is developed. From the developed BN models, risk probabilities and expected utilities (impacts) are deduced for on-board equipment in cooperative adaptive cruise control and signal controllers with and without redundant traffic surveillance systems. Impacts of propagated risks on CACC and signalized traffic network applications are quantified via simulations in average speed, average and maximum queue lengths, and delays which can result as incidents, energy loss, and higher emissions. For CACC application, impact levels result as $50\%$ delay difference when low amount of speed information is perturbed. When a significantly different speed characteristics is inserted by an attacker. Delay goes beyond $100\%$. At signalized networks with and without redundant systems, risk can increase average queues and delays by $3\%$ and $15\%$ and $4\%$ and $17\%$ respectively. However, further improvements are possible:
\begin{enumerate}[i]
\item Presented method can be used for attack modeling,
\item Expressing the systems as flow network for possible attack paths in order to optimize sensor deployment and minimize communication delays,
\item Alternative CV market penetration incorporation can be included, 
\item Developing detection models for possible attacks and mitigation efforts.  
 \end{enumerate}      
\section*{Acknowledgments}
This research is supported by U.S. Department of Homeland Security Summer Research Team Program for Minority Serving Institutions and follow-on grants managed by ORAU and was conducted at Critical Infrastructure Resilience Institute, Information Trust Institute, University of Illinois, Urbana-Champaign. The research also partially supported by U.S. Department of Transportation Regional Tier 1 University Transportation Center for Connected Multimodal Mobility. It is also partially supported by NSF Grants Nos. 1719501, 1436222, 1954532, and 1400991.

\bibliographystyle{elsarticle-harv}
\bibliography{ciri_trp}

\end{document}